%% file: ms.tex
\documentclass[twocolumn]{aastex62}

\defcitealias{2005ApJS..159..141V}{VF05}
\defcitealias{Brewer:2016gf}{B16}
\defcitealias{2017AJ....154..107P}{P17}

\newcommand{\teff}{$T_{\mathrm{eff}}$}
\newcommand{\logg}{$\log g$}
\newcommand{\monh}{[M/H]}
\newcommand{\kepler}{\textit{Kepler}}

\newcommand{\vbroad}{$v_{\rm broad}$}
\newcommand{\vsini}{$v \sin i$}
\newcommand{\vmac}{$v_{\rm mac}$}

\newcommand{\vrad}{$v_{\rm rad}$}

\newcommand{\inlinecode}{\texttt}

\newcommand\Nspecs{1,185}	
\newcommand\Nstars{1,149}      
\newcommand\Nearths{433}      
\newcommand\Nneptunes{267} 
\newcommand\Nplanets{1,562} 
\newcommand\Nplanethosts{1,003}    
\newcommand\Nduplicates{77} 
\newcommand\Ncommon{1,011} 

\newcommand\Nearthhosts{335}      
\newcommand\Nneptunehosts{216} 

\shorttitle{SPOCS Kepler Sample}
\shortauthors{Brewer \& Fischer}

\begin{document}

\title{Spectral Properties of Cool Stars: Extended Abundance Analysis of Kepler Objects of Interest}

\correspondingauthor{John M. Brewer}
\email{john.brewer@yale.edu}

\author[0000-0002-9873-1471]{John M. Brewer}
\affiliation{Department of Astronomy, Yale University, 52 Hillhouse Avenue, New Haven, CT 06511, USA}
\affiliation{Department of Astronomy, Columbia University, 550 West 120th Street, New York, New York 10027}
	\email{john.brewer@yale.edu}

\author[0000-0003-2221-0861]{Debra A. Fischer}
\affiliation{Department of Astronomy, Yale University, 52 Hillhouse Avenue, New Haven, CT 06511, USA}
	\email{debra.fischer@yale.edu}

\begin{abstract}

Accurate stellar parameters and precise elemental abundances are vital pieces to correctly characterize discovered planetary systems, better understand planet formation, and trace galactic chemical evolution.  We have performed a uniform spectroscopic analysis for 1127 stars, yielding accurate gravity, temperature, and projected rotational velocity in addition to precise abundances for 15 elements (C, N, O, Na, Mg, Al, Si, Ca, Ti, V, Cr, Mn, Fe, Ni, and Y).  Most of the stars in this sample are \kepler\ Objects of Interest, observed by the California-Kepler Survey (CKS) and include \Nplanethosts\ stars hosting \Nplanets\ confirmed planets.  This catalog extends the uniform analysis of our previous catalog, bringing the total of homogeneously analyzed stars to almost 2,700 F, G, and K dwarfs.  To ensure consistency between the catalogs, we performed an analysis of our ability to recover parameters as a function of S/N ratio and present individual uncertainties as well as functions to calculate uncertainties for parameters derived from lower S/N ratio spectra.  With the updated parameters, we used isochrone fitting to derived new radii, masses and ages for the stars.  We use our abundance analysis to support the finding that the radius gap is likely a result of evolution rather than the result of primordial compositional differences between the two populations.

\end{abstract}

\keywords{stars: solar-type}

%
\section{Introduction} \label{sec:intro}

Our understanding of planets and the planet formation process is almost entirely dependent on the photons collected from their host stars.  Planet mass and radius depend directly on the stellar mass and radius for planets detected through radial velocities and transits.  Broad categories of planet composition (rocky, icy, gaseous) are inferred from their average density \citep[e.g.][]{Zeng:2016cz,2009ApJ...693..722G,2007ApJ...669.1279S}, although more recently we have begun to use transmission and reflectance spectra of close in planets to directly probe the compositions of their atmospheres \citep[e.g.][]{2017AJ....154...95H,Morley:2016vc,2017MNRAS.469.1979M,Kreidberg:2015er}.

Despite the meager evidence in any given system, the rapid increase in planet discoveries allows us to exploit the demographics of exoplanet hosts to probe the planet formation process.  The planet metallicity correlation \citep{2005ApJ...622.1102F,2004A&A...415.1153S} provided strong support for the core-accretion model over gravitational stability by comparing systems with giant planets on short period orbits to stars without such planets.  As the number of discovered planets has exploded, we now know that almost all stars have planets \citep{2015ApJ...809....8B,2012ApJS..201...15H,2015ARA&A..53..409W}. With our current planet detection limits, it is impossible to construct a control sample of stars without planets. Instead, we have begun to compare the host metallicities of different types of planets, finding that smaller planets form around stars with a wider range of metallicities than hot Jupiters \citep{2014Natur.509..593B}, though possibly that metal rich stars more easily form planets of all sizes \citep{2016AJ....152..187M,2015AJ....149...14W}.

Studies of correlations between composition and planet formation have focused on overall metallicity, or [Fe/H] as a proxy for total mass in heavy elements. It is easier to measure due to the large number of atomic lines and is thought to trace the solid surface density in the protoplanetary disk, which may lead to more efficient planet formation.  However, ices and their rocky formation sites may have a greater influence on the rapid formation of planetary cores \citep{2011ApJ...738...97B,2009Icar..200..672D,Robinson:2006cr}.  During the main sequence lifetimes of sun-like stars, the abundances of heavy elements in the stellar photosphere decrease by $\lesssim 0.1$~dex and ratios between elements remain almost unchanged \citep{2017ApJ...840...99D,2009ARA&A..47..481A,2002JGRA..107.1442T}.  This makes them excellent proxies for the initial composition in the protoplanetary disk for systems where we already know the outcome of planet formation.

Recently, two distinct size populations of small planets were identified in the initial batch of \kepler\ Objects of Interest (KOIs) \citep{2017AJ....154..109F}.  There is a bimodal distribution in planet radius with a deficit of planets between $1.5 R_{\oplus}$ and $2.0 R_{\oplus}$.  The distribution was predicted to exist based on models of post-formation atmospheric  loss due to photo-evaporation and corroborating evidence from early radius measurements \citep{2013ApJ...776....2L,2013ApJ...775..105O,Lundkvist:2016gf}.  Although photo-evaporation largely explains the radius distribution, differences in the compositions of the planetary cores could also play a role. \citet{2017ApJ...845...61U} show that varying the magnesium to silicon ratio effects the oxygen uptake in planetary cores, changing the mass and radius of the final planet.

The clearly defined planet radii provided in \citet{2017AJ....154..109F} provides a sample where we can explore differences in composition that may influence the planet formation process.  By examining the Mg/Si ratio of the small radius planet hosts, we can probe the initial compositions that went into forming their planetary cores.  

This work derives detailed elemental abundances for 15 elements for all of the stars in the CKS catalog \citep[][hereafter P17]{2017AJ....154..107P}, placing them on the same scale as the catalog of \citet[][hereafter B16]{Brewer:2016gf}.  In addition, we add additional observations of a handful of KOIs that had low S/N or were not included in the CKS catalog.  Because most of the CKS spectra were observed at lower S/N than those in the \citetalias{Brewer:2016gf} catalog, we derived updated uncertainties for the stellar parameters as a function of S/N.  We then examine the host compositions of the bimodal small planets identified in \citet{2017AJ....154..109F} to evaluate the possibility that there is a chemical difference that leads to the differing planet sizes.

%
\section{Data and Analysis}

The CKS program obtained Keck HIRES spectra of a large number of Kepler objects of interest (KOI) using the same instrumental setup as that in \citetalias{Brewer:2016gf}. This allowed us to easily apply the same analysis method to derive stellar properties and abundances for all of the stars. The CKS program targeted the first large data release of KOIs \citep{2011ApJ...736...19B}, expanded that to include all KOIs with $K_p < 14.2$, then included some more interesting fainter systems. For the magnitude limited sample, the mean S/N per pixel is 45 measured on blaze near 5500 \AA, but the fainter stars were observed at lower S/N.  The instrument setup for all stars uses the red collimator of HIRES with the C2 and B5 deckers, which both have the same width and a resolution of $\sim 62,000$. Before our analysis, the spectra are first continuum normalized using the automated procedure of \citet{2005ApJS..159..141V}, which uses an iterative polynomial fitting procedure that replaces deep lines with pixels from neighboring orders.  Residual trends in the continuum had no noticeable impact on the final spectroscopic parameters \citepalias{Brewer:2016gf}.

We used Spectroscopy Made Easy (SME) \citep{2017A&A...597A..16P} with a single line list in an iterative procedure to fit the global stellar parameters (\teff, \logg, [M/H], and total rotational broadening from \vmac\ and \vsini\ in a single parameter \vbroad) and abundances for 15 elements.  Initial model parameters were set to the solar abundance pattern \citep{Grevesse:2007cx} with temperature and gravity derived from the stellar color.    We fit first for the global stellar parameters and individual abundances of the $\alpha$-elements calcium, silicon, and titanium to account for deviations from a solar abundance pattern.  After the initial fit, the model parameters were used as the starting parameters with the temperature perturbed by $\pm 100$~K and the spectra were re-fit.  Using the average parameter values of the three fits, the global parameters were fixed and we fit for the abundances of 15 elements (C, N, O, Na, Mg, Al, Si, Ca, Ti, V, Cr, Mn, Fe, Ni, and Y).  With this new abundance pattern, we iterate the procedure once.  Finally, we fix the macroturbulence using the relation derived in \citetalias{Brewer:2016gf} and fit for the projected rotational velocity (\vsini).

\subsection{Uncertainties and Signal-to-Noise} \label{sec:snr_uncertainties}

The majority of the spectra in \citet{Brewer:2016gf} were observed at a S/N per pixel $> 100$ and the uncertainties for that analysis were derived from that data.  The authors noted that the scatter in the recovered abundances doubled for the spectra with S/N $< 80$. Because the targets in the CKS sample are fainter, the mean S/N is $\sim 50$ with a significant fraction having S/N~$< 40$.  To quantify the impact of decreasing S/N on our ability to reliably recover stellar parameters and abundances across a variety of stellar types, we simulated observations of the Sun and a cool star at a range of S/N ratios and analyzed those spectra as if they were real observations using the procedure above.  The increase in the standard deviation in the returned parameters over that at S/N~$= 100$ for each of the S/N ratios then gives us a measure of the increased uncertainty for the lower S/N ratio spectra.

To generate the simulated observations of the Sun, we convolved the solar atlas of \citet{2011ApJS..195....6W} with a Gaussian of the same width as the instrumental profile of the Keck HIRES B5 decker using the \inlinecode{gaussbroad} function from SME.  We then resampled the spectrum to match the pixel scale from HIRES using the \inlinecode{resamp} function, also from SME.  With this `observation' we then generated 25 realizations of the spectrum at S/N ratios of 20, 30, 40, 50, 60, 80, and 100 by adding gaussian random noise to every pixel.  For the cool star, we chose an existing spectrum with a S/N $\sim 380$ and added noise to generate 25 new spectra in the same bins as for the simulated solar spectra.  After analyzing these simulations, we found the standard deviation in all parameters for the two sets of 25 spectra as a function of S/N to quantify the changing uncertainty.

\subsection{Small Planet Hosts}

Using the catagories of small planets identified by \citet{2017AJ....154..109F}, we selected all planets with $R_p <$ 1.8 $R_{\oplus}$ as `Super-Earths' and all with $1.8 < R_{\oplus} < 3.5$ as `sub-Neptunes'. \citet{2017AJ....154..109F} also made several quality cuts in the data to avoid contaminating their sample with stars with poorly determined parameters.  Specifically, they removed false positives, stars fainter than $K_p < 14.2$, transit impact parameters greater than 0.7, orbital periods greater than 100 days, evolved stars, and stars outside of the range $4700~\mathrm{K} < T_{\mathrm{eff}} < 6500~\mathrm{K}$. We performed the same cuts to remain consistent with their analysis, though a check on the full sample without those cuts showed no qualitative difference in our results.  This resulted in \Nearths\ super-Earth sized planets and \Nneptunes\ sub-Neptunes.  Most stars host only planets that are all about the same size \citep{2018AJ....155...48W}, and so we are tempted to remove from our sample those outlier hosts that contain both super-Earths and sub-Neptunes ($\sim 10\%$).  However, we could not guarantee that our sample would be free of mixed systems, only that we had not detected them.  We allow these hosts to count in both samples when comparing hosts instead of planets.  Finally, we excluded any planets that were around evolved stars to avoid both evolutionary effects and analysis trends in our abundances.  The resulting sample contained \Nearths\ Earth sized planets around \Nearthhosts\ dwarf hosts, and \Nneptunes\ sub-Neptunes around \Nneptunehosts\ dwarf hosts.

%
\section{Results} \label{sec:results}

We performed a homogeneous analysis of planet search stars, most of which are confirmed planet hosts, to derive global stellar parameters and abundances for 15 elements (C, N, O, Na, Mg, Al, Si, Ca, Ti, V, Cr, Mn, Fe, Ni, and Y) using the procedure of \citetalias{Brewer:2016gf}.  We made cuts identical to those in \citetalias{Brewer:2016gf} to keep only stars in the temperature ($4700~\mathrm{K} \leq T_{\mathrm{eff}} \leq 6800~\mathrm{K}$), gravity ($2.5 \leq \log{g} \leq 5.0$), metallicity ($-2.0 \leq [M/H] \leq 0.6$), and rotation ($0.5~\mathrm{km/s} \leq v_{\mathrm{rot}} \leq 25.0~\mathrm{km/s}$) ranges where our procedure returns consistent results.   The final set of parameters comes from \Nspecs\ spectra of \Nstars\ stars, mostly \kepler\ Objects of Interest (KOIs).  Of those, \Nduplicates\ duplicate stars in the previous catalog are included here to look at abundance differences in small planet hosts.  Due to the low average S/N ratio of the spectra, we also simulated spectra at a range of S/N ratios to derive uncertainties for the spectra.  Finally, we used a subset of the stars, defined in \citet{2017AJ....154..109F} to look for correlations between planet radius and stellar composition.  We break out each of these results below.

\begin{figure*}[htb!] 
   \centering
   \includegraphics[width=\textwidth]{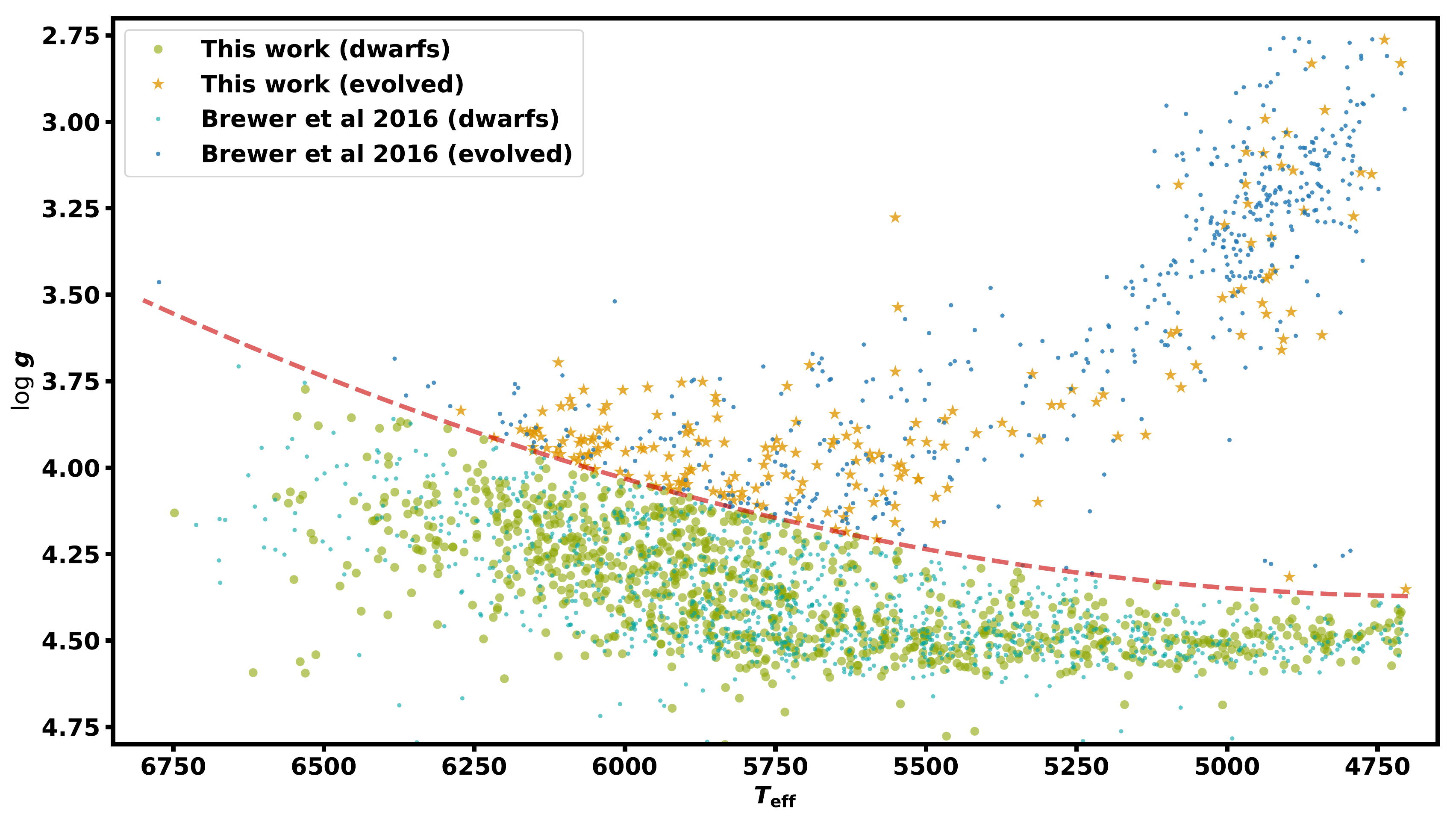} 
   \caption{All of the stars in this sample (green filled circles, gold filled stars), overlaid on those of \citetalias{Brewer:2016gf} (light blue and dark blue points), which used the same analysis technique.  Due to both evolutionary effects and the increasing influence of NLTE for evolved stars, we have divided the sample into dwarfs (green and light blue) and evolved (gold and dark blue) indicated by the red dashed line.  The majority of the stars in this work are \kepler\ Objects of Interest (KOIs) observed by the California-Kepler Survey (CKS) and are dwarfs.  The point colors correspond to the same stellar groups for other plots in this work.}
   \label{fig:kiel_diagram}
\end{figure*}

\subsection{Spectroscopic Stellar Properties}
We present the stellar parameters determined from fitting observed spectra with synthetic spectra in two tables.  Tables \ref{table:stellar_properties} and \ref{table:stellar_abunds} match Tables 8 and 9 in \citetalias{Brewer:2016gf} and contain the global stellar parameters and the final trend-corrected abundances respectively.  This layout was chosen to better facilitate combining the two data sets. In Table \ref{table:uncertainties} we provide the per-star uncertainties for both the global stellar parameters and abundances using Equation \ref{eqn:new_sigma}.

The SPOCS ID is in Column (1) of Tables \ref{table:stellar_properties} and \ref{table:stellar_abunds}, followed by the star name in column (2).  The SPOCS ID numbers are shared with \citet{2005ApJS..159..141V} and \citetalias{Brewer:2016gf} and all stars new to this catalog have ID numbers $> 3247$.  Columns (3) and (4) of Table \ref{table:stellar_properties} contain the RA and DEC of the star, Columns (5)-(7) contain the \teff, \logg, and \monh\ derived from this analysis.  Note that \monh\ is the metallicity of the model atmosphere used to derive the best fit model spectrum but does not necessarily match the [Fe/H] value for the star.  For stars with available broadband colors and $0.4 < B-V < 1.0$, we calculated both of the chromospheric activity indicators $S_{HK}$ (Column (8)) and $\log R^{'}_{HK}$ (Column (9)) values following the procedure outlined in \citet{Isaacson:2010gk}.  In modeling the spectra, we first derive the total rotational broadening, \vbroad\ (Column (10)) then using the final parameters and abundances fix \vmac\ (Column (12)) using the relations in \citetalias{Brewer:2016gf} and solve for \vsini\ (Column (11)).  The barycentric radial velocity, \vrad\ (Column (13)), is the average over all epochs.  The average SNR (Column (14)) for the spectra along with the average RMS residuals between model and observation in the continuum and line regions (Columns (15) and (16)) give a measure of the fit.  The final column of Table \ref{table:stellar_properties} (Column (17)) contains the number of spectra analyzed that were averaged to give the final parameters and abundances.

The stars in the sample span dwarfs to early giants or late subgiants from 4700~K to 6800~K (Figure \ref{fig:kiel_diagram}).  As a star evolves off the main sequence, abundances in the photosphere can be altered by changes in the extent of the convective zone.  When making abundance comparisons, we will limit ourselves to stars on the main sequence, and so divide our sample using the \teff\ and \logg\ relation developed in Myles et al (2018 in prep).

\subsection{Elemental Abundances}
\begin{figure*}[htb!] 
   \centering
   \includegraphics[width=\textwidth]{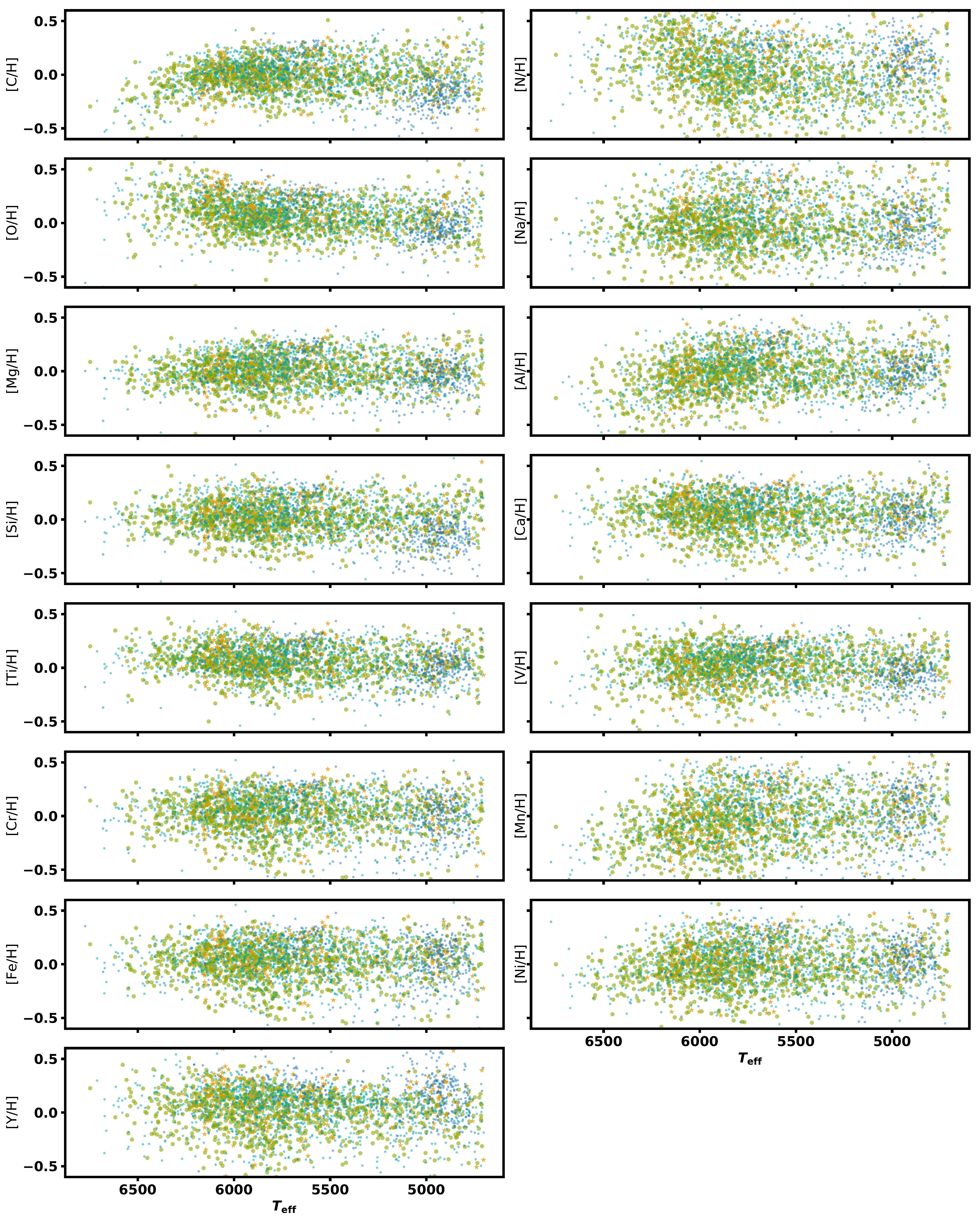} 
   \caption{Elemental abundances for 15 elements (C, N, O, Na, Mg, Al, Si, Ca, Ti, V, Cr, Mn, Fe, Ni, and Y) plotted against temperature.  There are no significant trends in the dwarf abundances (green circles) and they match the distribution of dwarfs from \citetalias{Brewer:2016gf} (light blue points).  The evolved stars (gold stars from this work, dark blue points from \citetalias{Brewer:2016gf}) show an offset to higher metallicities that may be due to these stars being younger, on average, than longer lived dwarfs. For some elements, NLTE effects can induce trends in abundances with temperature and gravity.  In this analysis, oxygen and manganese are the most affected, particularly for hotter and more evolved stars, and these trends can be seen here.}
   \label{fig:elems_x_h}
\end{figure*}

In Table \ref{table:stellar_abunds}, Columns (18)-(32) contain the solar relative log number abundances for 15 elements.  The abundances are differential with respect to the Sun, using asteroid spectra as a reference, and then calibrated to minimize trends in abundance with temperature using a sample of high SNR dwarf spectra \citep{Brewer:2016gf}.  The abundances for both main sequence and evolved stars cover the same ranges as those in \citetalias{Brewer:2016gf} (Figure \ref{fig:elems_x_h}).  For most elements there are no strong trends in abundance with \teff\ for dwarfs. However, for elements that suffer from NLTE effects, such as oxygen and manganese at temperatures above 6100~K, evolved stars are more strongly affected. This behavior is expected due to the sensitivity of NLTE effects to atmospheric structure \citep{2012MNRAS.427...27B} and care should be taken when comparing abundance ratios of evolved and main sequence stars.

\subsection{Comparison to CKS}
The CKS Catalog includes parameters for 1305 stars including some that are hotter or cooler than those included in this analysis.  In addition, some spectra analyzed by \citetalias{2017AJ....154..107P} did not successfully converge in our analysis due to low S/N or did not meet our quality cuts for fit, $4700~\mathrm{K} <$ \teff\ $< 6800~\mathrm{K}$, $2.5 <$ \logg\ $< 5.0$, or $0.5~\mathrm{km/s} <$ \vbroad\ $< 25$~km/s.  In total, there are \Ncommon\ stars in common between the two catalogs.  We discard one of those (KOI-1060) in the following discussion because \citetalias{2017AJ....154..107P} report its \vsini\ as 456~km/s, which is likely a failure in their interpolation for this star as visual inspection shows the spectrum to be a low S/N but slowly rotating star (\vsini\ $= 7.7$ km/s) and is well fit by our model.

\begin{figure}[htbp!]
	\includegraphics[width=\columnwidth]{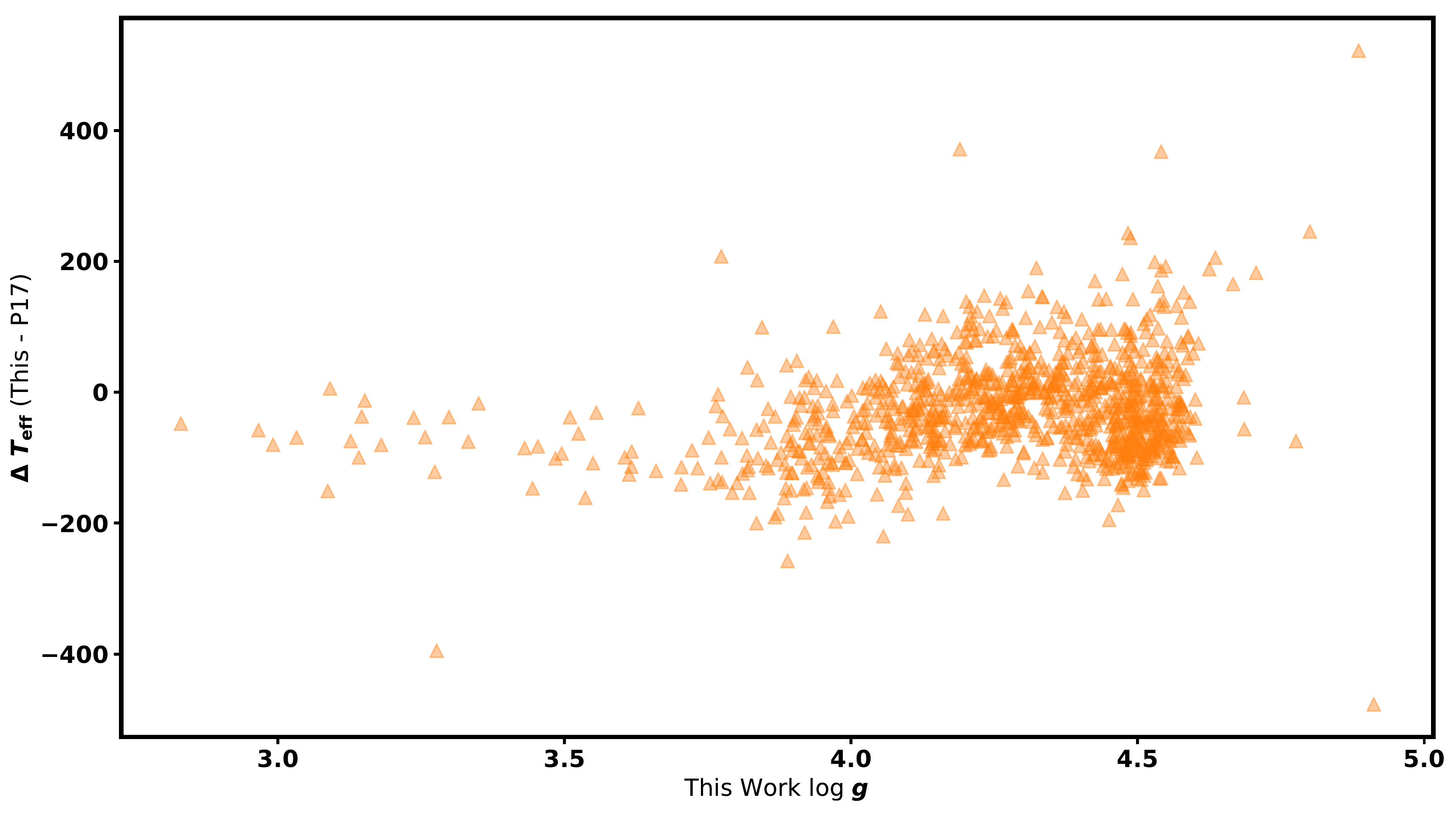}
	\caption{For main sequence stars there appears to be a small offset between \citetalias{2017AJ....154..107P} and this work.  Stars with $3.7 < \log g < 4.4$ also show a slight trend of decreasing \teff\ with decreasing \logg.}
	\label{fig:teff_trend}
\end{figure}

\begin{figure*}[htbp!] 
   \centering
   \includegraphics[width=\textwidth]{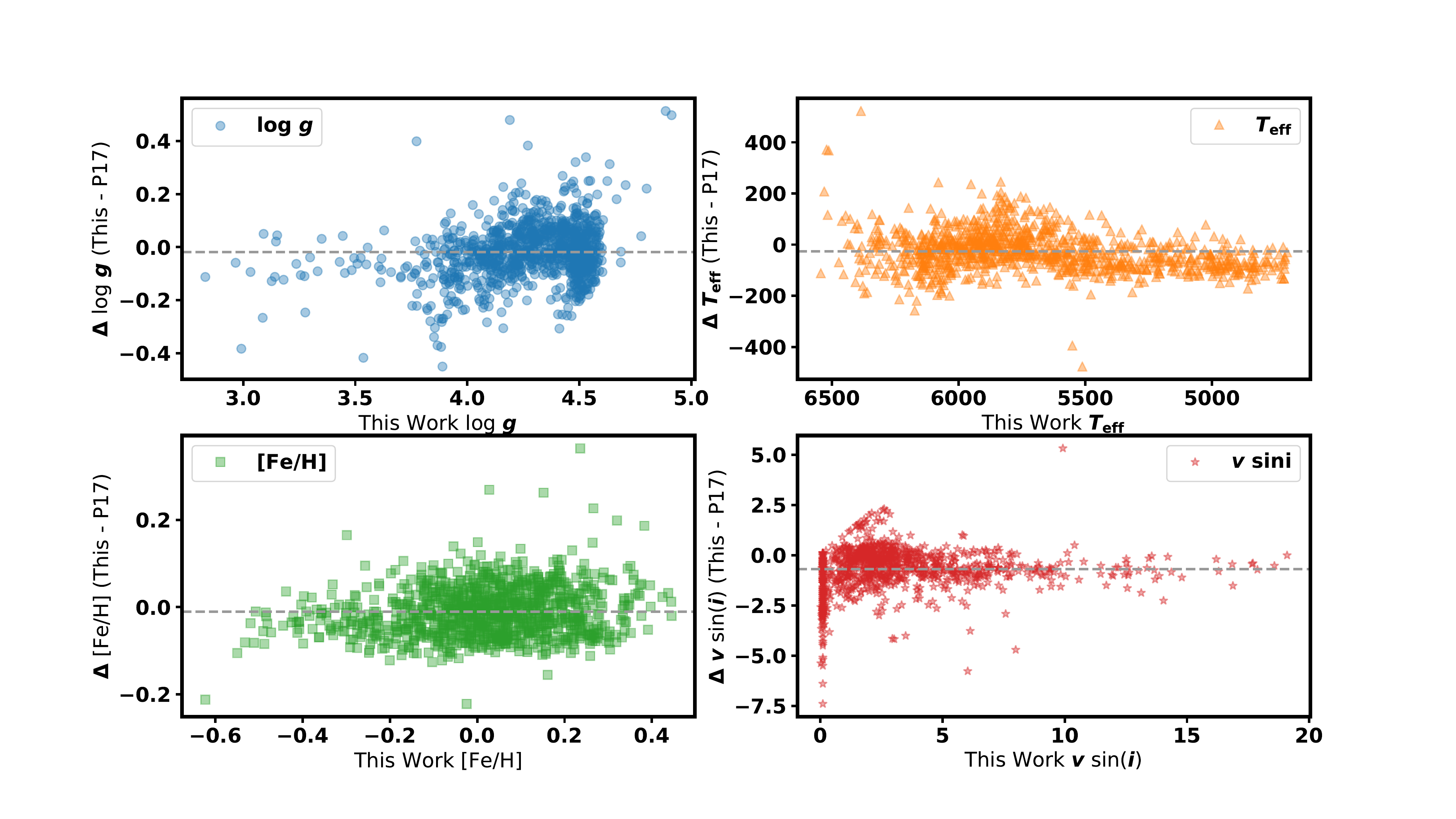} 
   \caption{Differences between the parameters derived in this work and those from \citetalias{2017AJ....154..107P} with the means plotted as dashed grey lines.  The CKS analysis is based on interpolation within a grid of models or observed spectra and has larger uncertainties than those in this work. For most of the 1010 stars in common, all parameters agree to within the combined uncertainties, however effective temperatures have slightly higher differences that are likely due to the 250~K spacing of their model grid.}
   \label{fig:compare_cks}
\end{figure*}

In general, the parameters in the CKS catalog and this work agree to within our mutual uncertainties, although there is a slightly higher dispersion in \teff\ (Figure \ref{fig:compare_cks}).  There is a 26~K offset in the mean \teff\ between the two samples, and the standard deviation about this mean is 79~K.  When comparing the difference in temperatures as a function of \logg\ there is a trend between $3.7 \lesssim \log g \lesssim 4.4$ (Figure \ref{fig:teff_trend}).  At lower gravities as measured in this work, the temperatures from \citetalias{2017AJ....154..107P} are systematically higher, decreasing to be equal at about solar \logg, then increasing sharply again at higher gravities.  This behavior can be seen in the temperatures alone in the \teff\ panel of Figure \ref{fig:compare_cks} and there is a slight trend in surface gravity itself at $\log g > 3.7$.  As these issues were not encountered in comparisons with other catalogs in \citetalias{Brewer:2016gf}, we assume that these are an artifact of the interpolation process in \citetalias{2017AJ....154..107P}.  The temperature average was forced to agree with our scale, however the fluctuations match roughly with the 250~K spacing of their model grid (Petigura, private communication).

\subsection{Uncertainties} \label{sec:results_uncertainties}
The expected outcome from adding increasing amounts of noise to our model is increasing uncertainty in our fit and should follow a 1/(S/N) relation.  The scatter in the fits to simulated solar and cool star spectra showed slightly different responses to the decreasing S/N ratio.  In general, the result followed a 1/(S/N) relation, with some parameters having a slightly flatter and others a slightly steeper relation.  Since the differences were small and inconsistent between the solar and cool star case, we chose the single functional form proportional to 1/(S/N) for all of the parameters.

\begin{figure*}[htb!] 
   \centering
   \includegraphics[width=\textwidth]{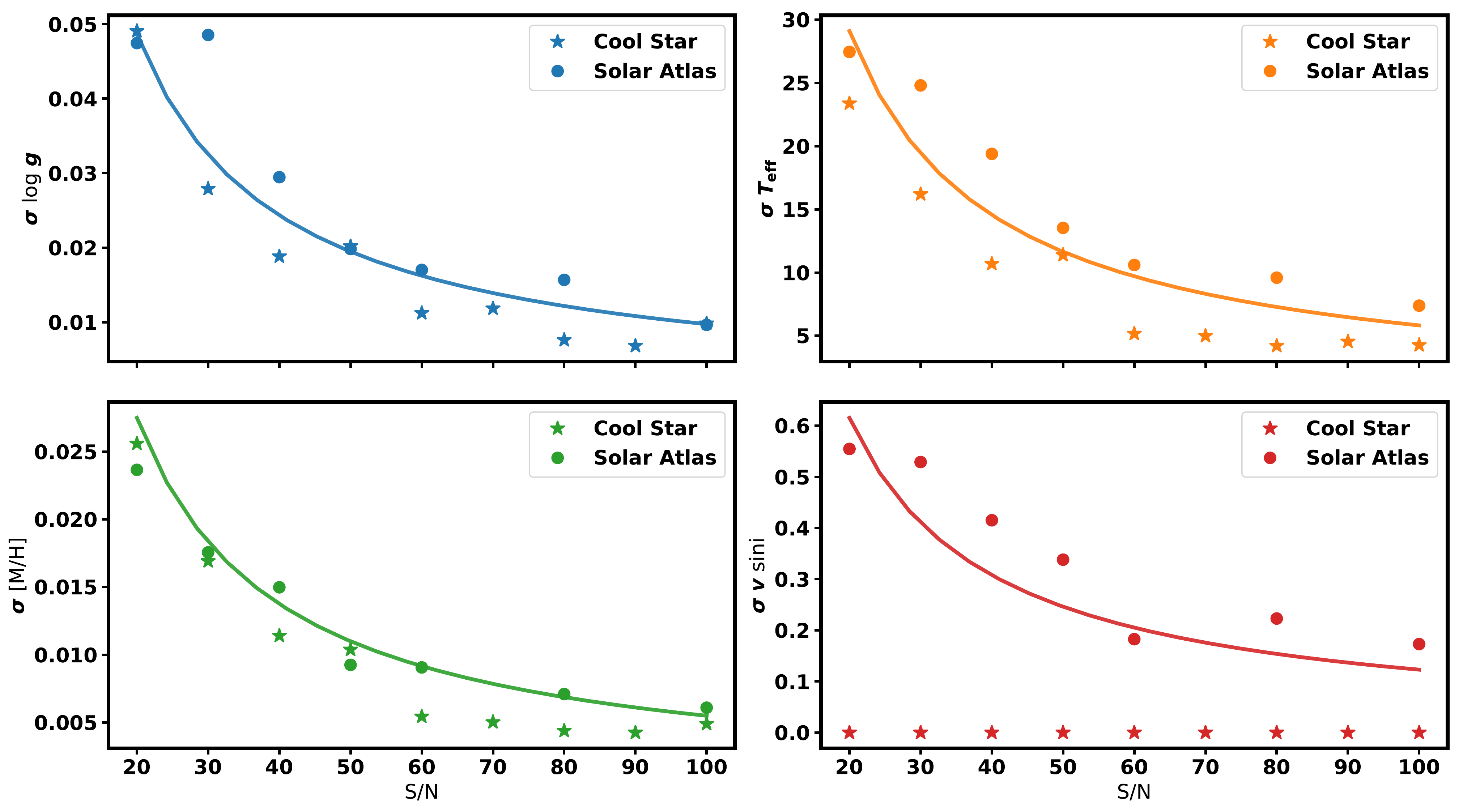} 
   \caption{Standard deviation in mean of returned parameters when analyzing 25 simulated spectra  of the Sun and 25 spectra of a cool star at varying S/N ratios.  The solid lines represent (1 + 1/(S/N)) times the standard deviation at S/N = 100.  We use this as an additional uncertainty term, added in quadrature to that derived in \citetalias{Brewer:2016gf}, for all spectra with S/N $< 100$.  Parameters for stars with S/N $\gtrsim 50$ are generally very precise and corrections to the uncertainties are small.  The cool star had negligible \vsini\ and the always returned the same value.}
   \label{fig:snr_param_sigmas}
\end{figure*}

\input{table_01.tex}

The increase in the scatter in returned parameters was approximately proportional to $\frac{1}{S/N}$, with some variations between the solar case and the cool star case.   To calculate the additional uncertainties for spectra with S/N $< 100$, we chose this form and began at the mean standard deviation at S/N $= 100$ of the Sun and cool star for each of the parameters (Equation \ref{eqn:stddev}).  The exception was \vsini, which was negligible in the cool star and had no variation in recovered \vsini. We simply subtracted 0.05 km/s from the solar case to obtain a reasonable approximation to the \vsini\ values.

\begin{eqnarray}
	\Delta \sigma = \sigma_{100} * \frac{100}{S/N} \label{eqn:stddev}
\end{eqnarray}

To calculate the additional uncertainties for spectra with S/N $< 100$, we then calculated this additional $\sigma$ for all of the parameters and abundances using Equation \ref{eqn:stddev}, where $\sigma_{100}$ is the mean standard deviation between the solar and cool star case at S/N $= 100$, and added them in quadrature to the uncertainties provided in \citetalias{Brewer:2016gf}  (Equation \ref{eqn:new_sigma}). The uncertainties in that work were calculated by determining the distribution of differences between all stars with multiple observations and S/N $> 100$, then multiplying by two to account for possible systematic trends. Table \ref{table:fit_coeffs} provides the means for Equation \ref{eqn:stddev} to calculate new uncertainties (Equation \ref{eqn:new_sigma}) for the lower S/N spectra from \citet{Brewer:2016gf}.

Two parameters that differed significantly between the solar and cool star simulations were \vsini\ and [N/H].  In the case of \vsini, the cool star showed negligible rotational broadening and our model fitting consistently returned the minimum \vsini.  

\begin{eqnarray}
	\sigma_{new} & = & \sqrt{\sigma_{b16}^2 + \Delta\sigma^2} \label{eqn:new_sigma}
\end{eqnarray}

The uncertainties listed in Table \ref{table:uncertainties} are the statistical uncertainties from fitting models to observations.  They provide a measure of the precision but, with the exception of \logg, do not incorporate any estimate of the accuracy of the parameters.  At solar temperature, gravity, and metallicity \citetalias{Brewer:2016gf} demonstrated that the parameters and abundances were accurate, consistently returning the solar values.  

\begin{figure}[htbp!] 
   \centering
   \includegraphics[width=\columnwidth]{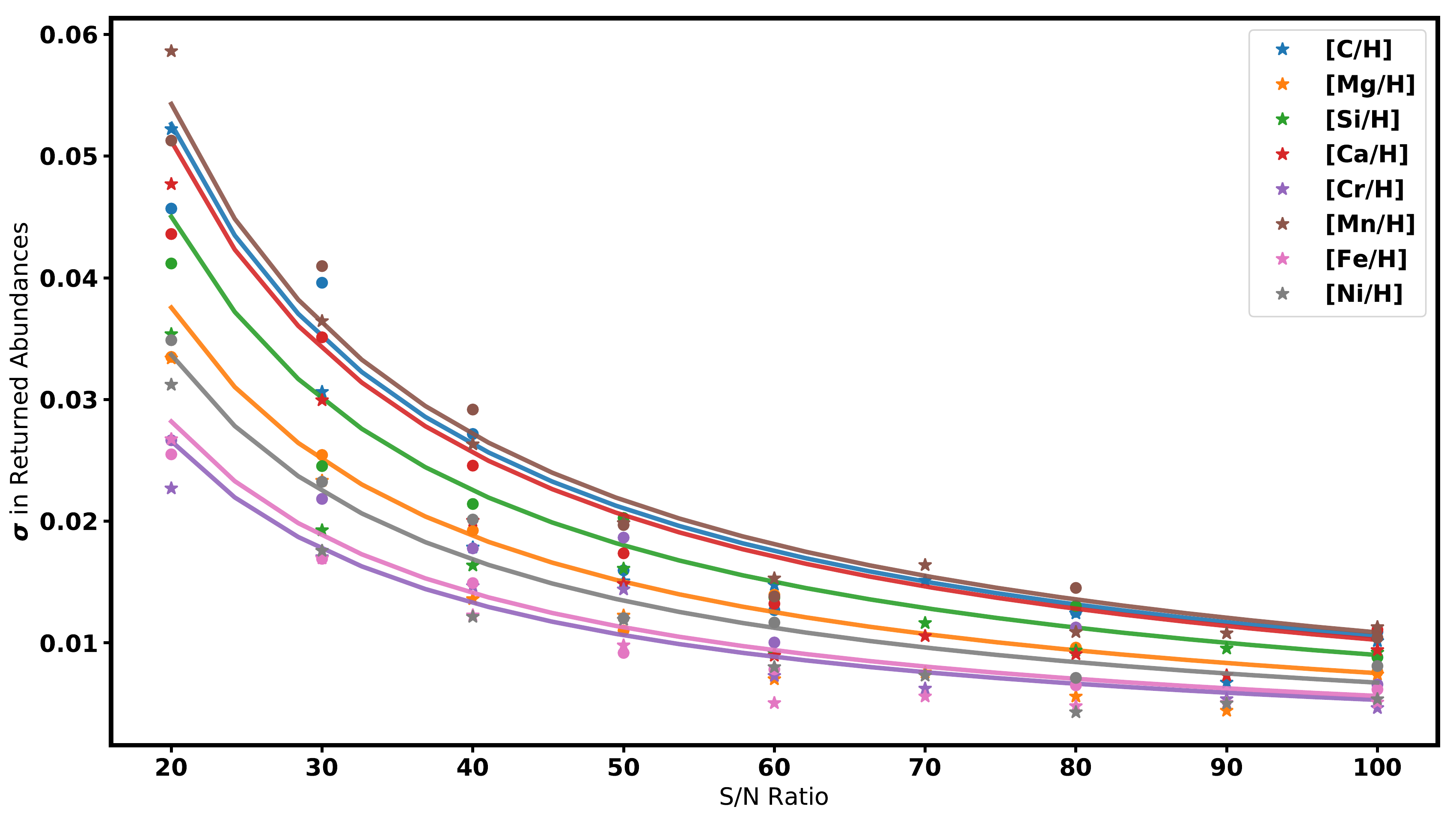} 
   \caption{Standard deviation in mean of returned abundances for elements that started with scatter of $\sim 0.01$ dex at a S/N of 100.  The solid lines are $\propto \frac{1}{S/N}$ and reasonably approximate the increased scatter in the abundances.  This is added in quadrature to the uncertainty derived in \citetalias{Brewer:2016gf}, for all spectra with S/N $< 100$.}
   \label{fig:snr_abunds_low_sigma}
\end{figure}

\begin{figure}[htbp!] 
   \centering
   \includegraphics[width=\columnwidth]{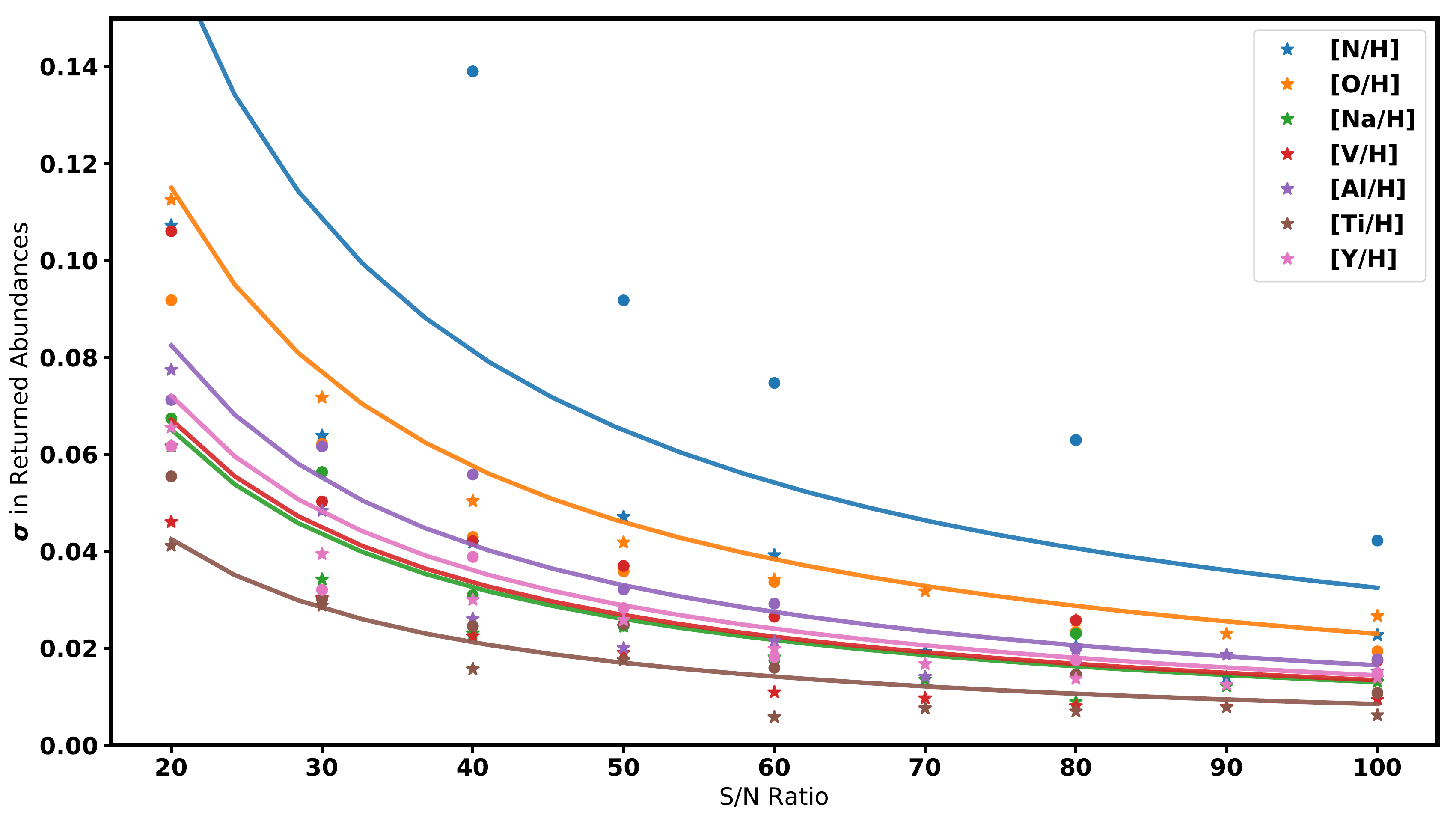} 
   \caption{Standard deviation in mean of returned abundances for elements that started with scatter of $\sim 0.02$ dex at a S/N of 100.  The solid lines are $\propto \frac{1}{S/N}$ and reasonably approximate the increased scatter in the abundances. This is added in quadrature to the uncertainty derived in \citetalias{Brewer:2016gf}, for all spectra with S/N $< 100$.  Nitrogen abundances in the solar case have a larger scatter even at higher S/N and rapidly deteriorate in precision as S/N decreases.  However, for the cool star simulations the scatter in nitrogen abundance increases at about the same rate as oxygen.}
   \label{fig:snr_abunds_high_sigma}
\end{figure}

We also looked at the change in the mean itself to see if any of the parameters or abundances had systematic trends induced by the decreasing signal-to-noise.  In general, there are no appreciable or consistent shifts in the mean as a function of the decreasing S/N.  However, the mean returned nitrogen abundance decreased to -0.04 dex at S/N $= 30$ and -0.09 dex at S/N $= 20$.  We have added these offsets in quadrature to the uncertainties in the nitrogen abundance at these S/N ratios.  The mean in the oxygen and manganese abundances also show trends, though they are much smaller, reaching positive and negative 0.024 dex respectively at  S/N $= 20$ and do not appreciably affect the uncertainties.

\input{table_02.tex}

\subsection{Star and Planet Radii}
\begin{figure}[htb!!] 
   \centering
   \includegraphics[width=\columnwidth]{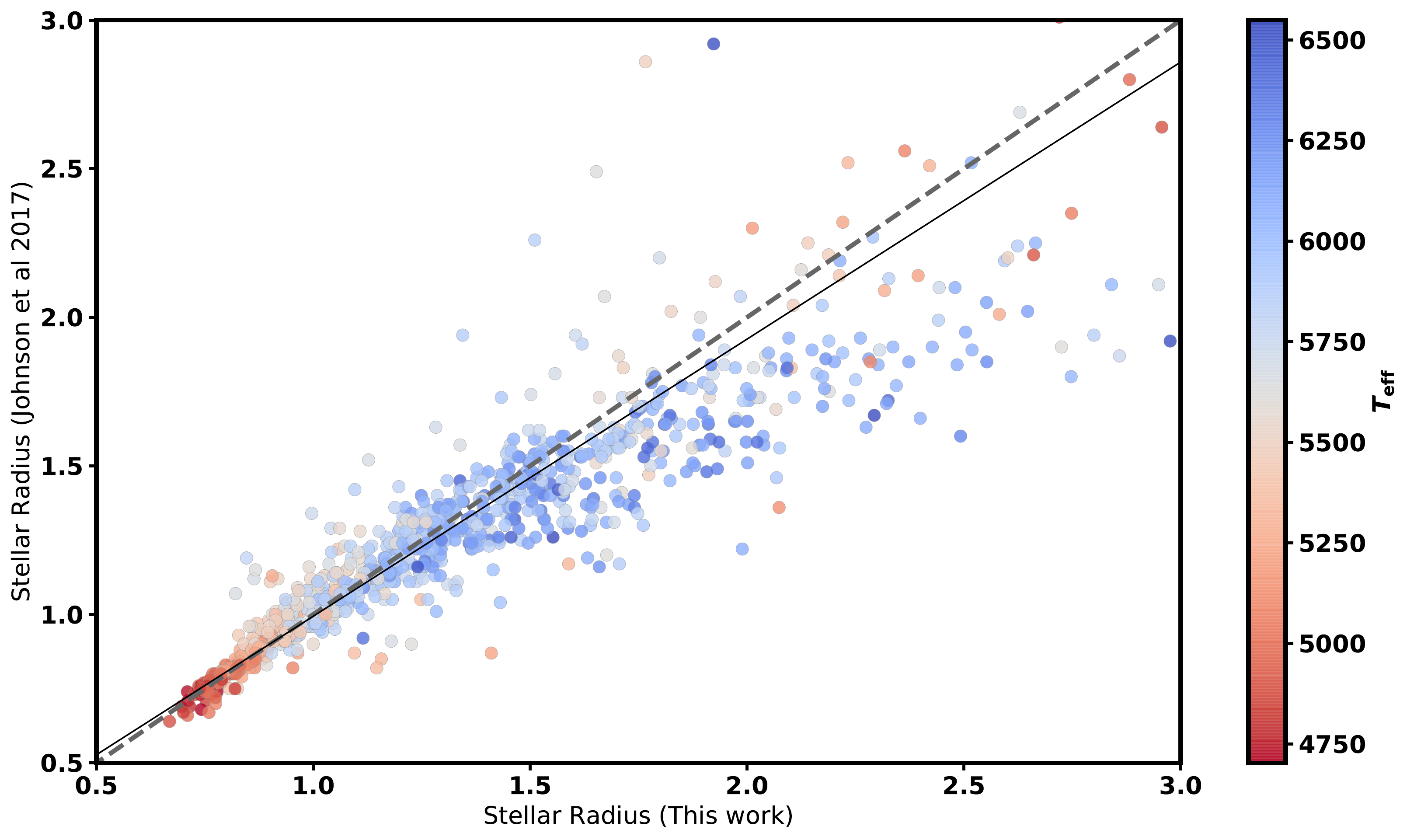} 
   \caption{A comparison of stellar radii from \citet{2017AJ....154..108J} compared to radii derived from isochrone fitting using our newly determined [Fe/H], \teff, and \logg\  in conjunction with Gaia DR2 parallaxes \citep{2016A&A...595A...1G}.  The grey dashed line shows the 1:1 line and the thin black line shows a simple linear fit to the uncertainty weighted mean of the sample.  The RMS scatter about this fit is $11\%$.}
   \label{fig:stellar_radius_compare}
\end{figure}
\begin{figure}[htb!!] 
   \centering
   \includegraphics[width=\columnwidth]{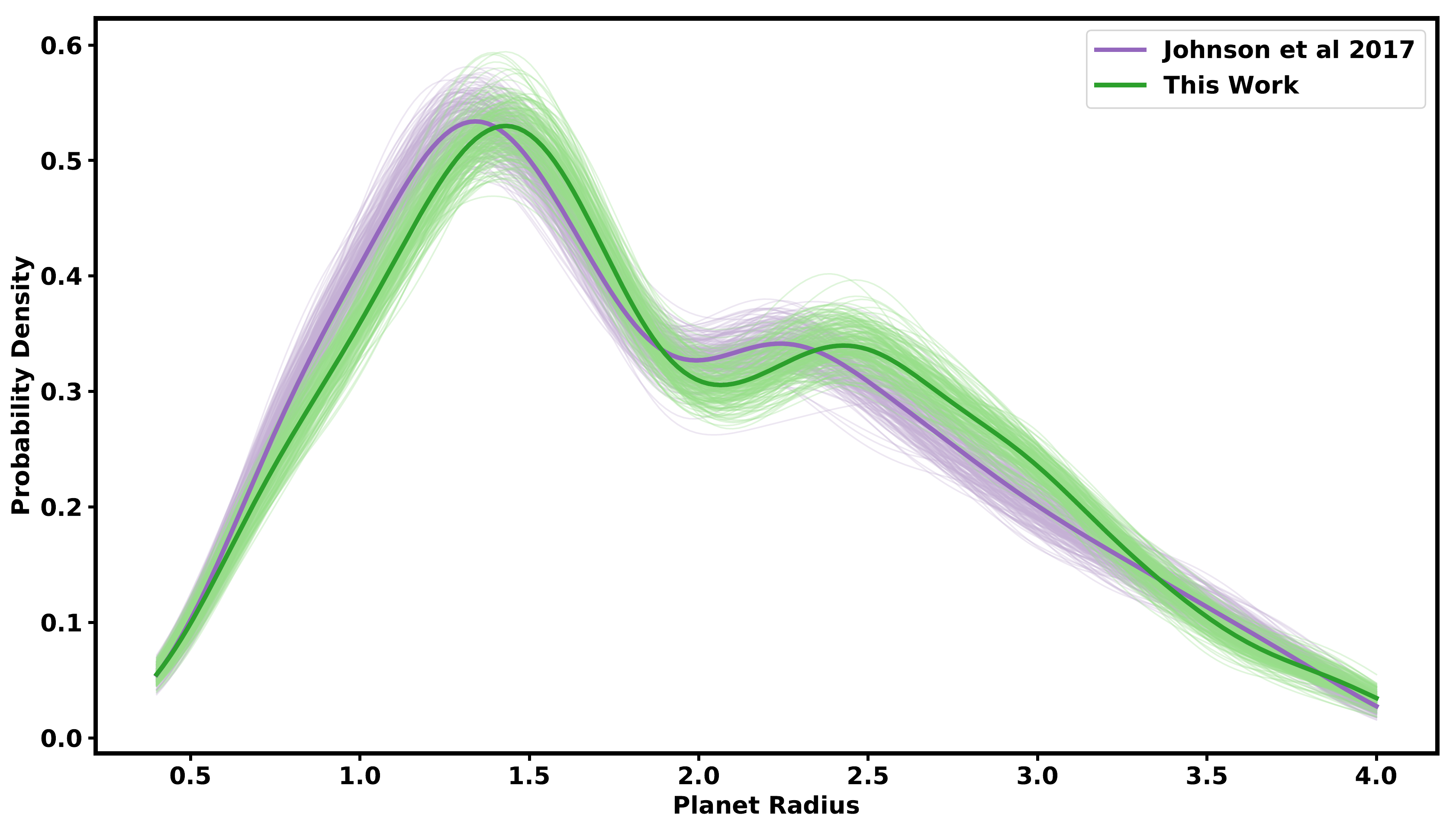} 
   \caption{Comparison of the probability density functions (PDFs) fit using a Gaussian kernel density estimate (KDE) of small planet radii derived in this work with those derived in \citet{2017AJ....154..108J}.  The solid heavy lines are fit to the actual radii and the light thin lines are fits to 200 bootstrap realizations of the data.  There is no significant difference between the two sets of radii, though our analysis seems to slightly favor larger radii.}
   \label{fig:planet_radius_compare}
\end{figure}

Although the differences between the parameters derived in the CKS catalog and this work were small, there were some systematic differences in both \logg\ and \teff.  Because the CKS values were then used to derive the stellar radius, and from that the planet radius \citep{2017AJ....154..108J}, these differences could have an effect on any interpretations related to the size of the planets.  We followed the same procedure as \citet{2017AJ....154..108J}, using the Python package \inlinecode{isochrones} \citep{2015ascl.soft03010M} to fit our parameters and 2MASS \citep{Skrutskie:2006hl} K magnitudes from SIMBAD \citep{Wenger:2000ef} to isochrone grids from the Dartmouth Stellar Evolution Database \citep{2008ApJS..178...89D}.  However, we also included Gaia DR2 parallaxes \citep{2016A&A...595A...1G}, which we obtained using the CDS cross-match service with a 1.5" radius, for all but 28 stars that were not found (5) or had no reported parallax (23). More than one Gaia source was found for about 30 of the stars, so we chose the closest match in these cases.    Additionally, five had negative parallaxes \citep{2018arXiv180409376L}, but this was still a useable constraint for the \inlinecode{isochrones} package. Seven stars without K magnitudes were not fit.  The isochrone fitting gives us radii, masses, and ages for the stars and we compile all of those along with their 1-$\sigma$ uncertainty ranges in Table \ref{table:derived_stellar_props}.  Including the parallax in the isochrone fits gave us slightly tighter constraints on the radius than we would have, giving us a median uncertainty in radius of only 4.4\%, with 90\% of the stars having an uncertainty of less than 20\%.

We found generally good agreement in the radii for most stars (Figure \ref{fig:stellar_radius_compare}), with an offset of $7.7\%$ and a weighted standard deviation of $11\%$, though there was a small systematic with larger stars having slightly larger radii in our analysis. The small planet comparison looked only at stars with smaller radii, where there is generally better agreement and smaller scatter.  We also re-derived the planet radii using our updated stellar radii and verified that the distribution of small planet radii was relatively unchanged with a small shift to larger radii (Figure \ref{fig:planet_radius_compare}).  We compared the full distributions of all matching planets between \citet{2017AJ....154..108J} and this work.  To estimate the uncertainties in the empirical PDFs, we generated 200 bootstrap realizations of the radii and found that they are consistent, though there may be a preference for slightly smaller radii of the sub-Neptunes with our new radii.

\subsection{Small Planet Compositions} \label{sec:results_planet_compositions}

\begin{figure}[htbp!] 
   \centering
   \includegraphics[width=\columnwidth]{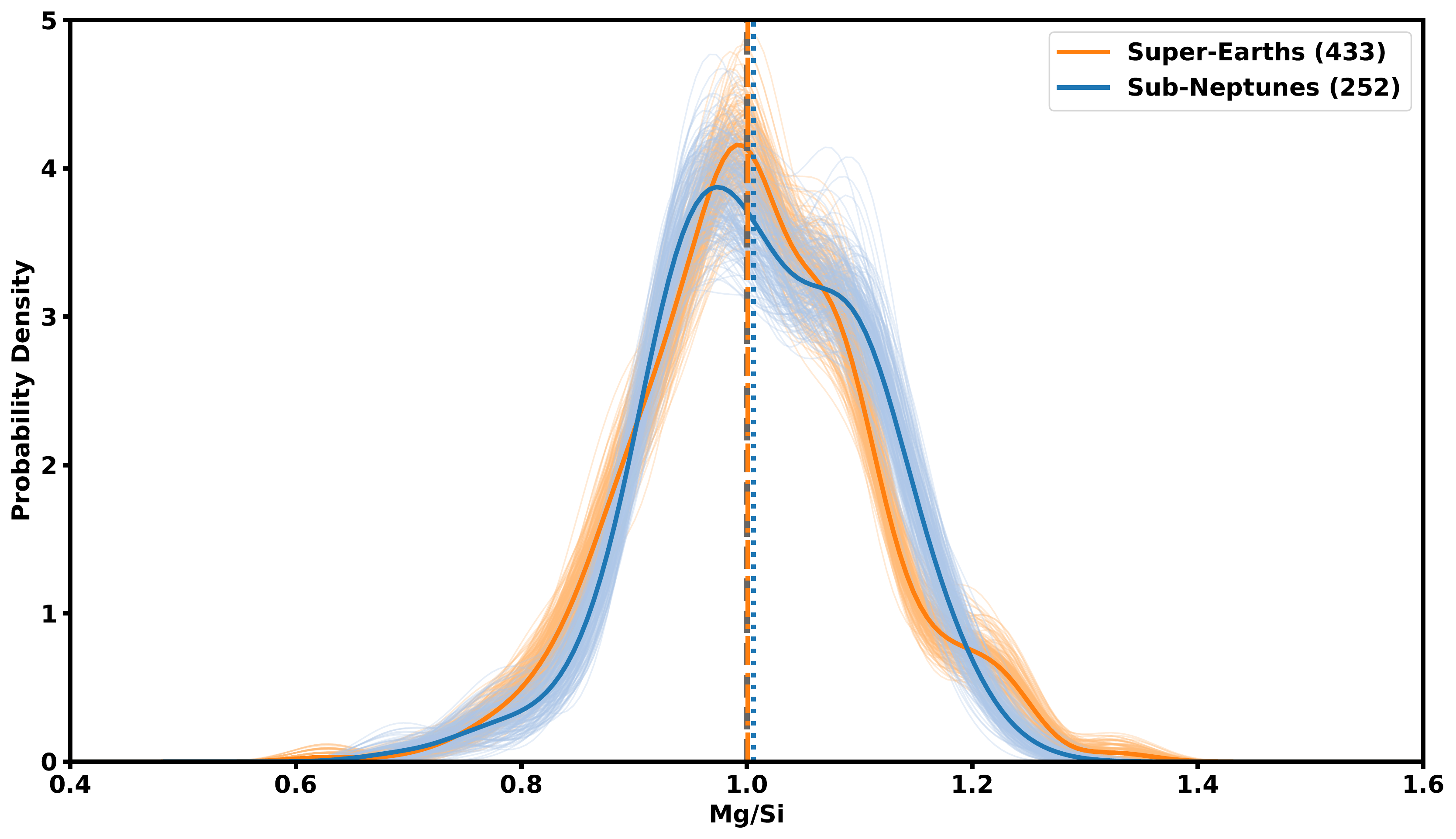} 
   \caption{The distribution of host star Mg/Si ratios for super-Earth sized planets (orange) is nearly identical to that of sub-Neptune sized planets.  Plotted are probability density functions (PDF) for the planets calculated using a Gaussian kernel density estimator.  The fainter lines of each color are PDFs generated in the same way from 200 bootstrap resamplings with replacement and the vertical lines mark the median of each distribution.  The PDFs are nearly identical and there does not appear to be any difference in the composition of these stars that might lead to differing planet radii, lending support to the idea of post-formation sculpting of the planets by photo-evaporation.  A similar comparison using only hosts and not planets, to avoid counting the same star twice, yields a qualitatively identical picture.}
   \label{fig:mgsi_pdf_hosts}
\end{figure}

Our hypothesis is that chemical differences in the protoplanetary disk, specifically in the Mg/Si ratio, could lead to differences in the mass of the planetary core.  This in turn could lead to a bimodal population of small planets.  To test this, we used a Gaussian kernel density estimator to calculate the probability density function (PDF) of the host Mg/Si distribution for the super-Earth and sub-Neptune sized  planets.

The Mg/Si distribution of planets with $R_p < 1.5 R_{\oplus}$ is statistically indistinguishable to that of planets with $1.5 R_{\oplus} < R_p < 3.5 R_{\oplus}$ (Figure \ref{fig:mgsi_pdf_hosts}). To estimate the uncertainties for the PDFs we used bootstrap resampling to generate 200 realizations of each sample and generated PDFs using a Gaussian kernel density estimator.  The PDFs appear very similar, but to quantify any differences between them we also performed a two sample Kolmogorov-Smirnov (KS) test on the cumulative distributions.  We found that the KS test could not reject the null hypothesis that both samples were drawn from the same distribution and returned a two sided p-value of 0.33.  Because the multiplicity of smaller planets is slightly higher than that of larger planets in our sample ($\sim 1.3$/star vs. $\sim 1.17$/star), we may be slightly shifting the distributions through weighting.  However, when we looked at just the distributions of hosts and not planets we saw a qualitatively identical distribution.  In fact, the host samples appear even more similar than when looking at the planets.  The KS test for the hosts returned a two sided p-value of 0.73.

\begin{figure}[htbp!] 
   \centering
   \includegraphics[width=\columnwidth]{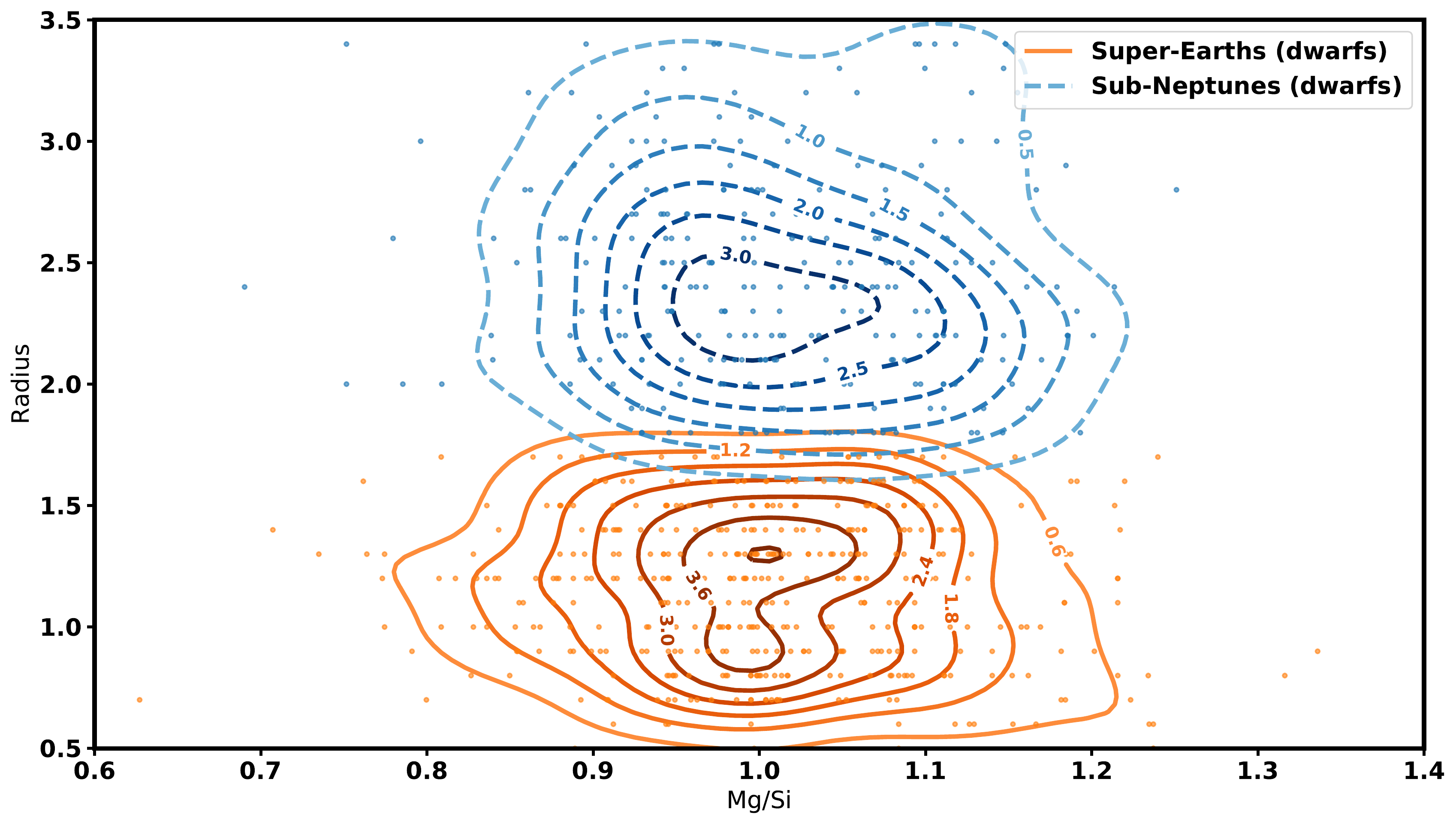} 
   \caption{Planet radius as a function of host star Mg/Si ratio for planets around dwarf stars with super-Earth and sub-Neptune planets.  The contours represent the two dimensional PDFs calculated using a Gaussian kernel density estimator as in the one-dimensional PDFs from Figure \ref{fig:mgsi_pdf_hosts}.  There is no apparent correlation between the Mg/Si ratio of a planet's host star and the radius of the planet in either the super-Earth or sub-Neptune samples.}
   \label{fig:mgsi_contours_planets}
\end{figure}

We also looked at the distribution of planet radius as a function of the Mg/Si ratio of their host stars by calculating a two-dimensional PDF in radius and Mg/Si space (Figure \ref{fig:mgsi_contours_planets}).  We found again that there was no appreciable difference in the distributions for super-Earth sized planets and sub-Neptunes.

%
\section{Discussion} \label{sec:discussion}

Stellar elemental abundances give us a window to the primordial compositions present in the protoplanetary disk.  Differences in composition can lead to differences in planet mass, radius, and geology as well as formation timescales by altering the availability of grains where ices form \citep{2016ApJ...831L..19O,2016ApJ...833..203P,2017ApJ...845...61U}.  The current state of the art in spectral analysis leads to very precise abundances, but comparisons between different analyses often far exceed the reported uncertainties \citep{2016ApJS..226....4H}.  To minimize the effects of varying analyses, we performed a uniform differential analysis of a large number of \kepler\ objects of interest, placing them on the same scale as that of \citetalias{Brewer:2016gf}.  This homogeneous catalog of almost 2,700 stars enables detailed relative abundance studies and can lead to insights into the planet formation process.

\subsection{Uncertainties} \label{sec:discussion_uncertainties}

One issue that can still affect abundance comparisons using this catalog is the varying S/N ratio of the spectra, especially for fainter stars that are generally observed at lower S/N ratios.  By simulating more than 700 observations at varying S/N, we showed that our analysis method was robust to the increasing noise.  At S/N ratios above 50, the RMS scatter in the returned parameters remained relatively small and there were no trends in the mean of the returned parameters and abundances.  Below a S/N of 50 the scatter in all parameters increased rapidly and in a few cases, particularly nitrogen abundance, the mean of the abundance also shifted away from the true value.  To facilitate comparisons for stars within the catalog, we quantified the increase in our uncertainty due to decreasing S/N and provide a table of individual uncertainties for the parameters of the \Nstars\ stars in this catalog.

The uncertainty corrections that we derived also apply to the previously analyzed stars in \citetalias{Brewer:2016gf}.  Though parameters for most of the stars in that catalog were derived using spectra with S/N $> 100$, almost 25\% used lower quality spectra.  The uncertainty equations here will allow their inclusion in abundance studies without risk of skewing the interpretation.

\subsection{Abundances} \label{sec:discussion_abundances}

The analysis procedure used in this work  has already been shown to produce accurate gravities in agreement with asteroseismic values to within 0.05~dex~\citep{2015ApJ...805..126B}, and other stellar parameters along with precise abundances that, for most elements, don't have strong trends with temperature \citep{Brewer:2016gf}.  In Figure \ref{fig:elems_x_h}, we can see that the stars added in this catalog follow the same general distributions, for both dwarfs and evolved stars, as those from the previous catalog.  One interesting feature that can be seen in the abundances is a slight difference in the distribution of dwarfs and evolved stars.  This could be due to either flaws in the analysis, which may have trouble at the lower gravities of evolved stars, or actual astrophysical differences at a given \teff, or both.

An evolved star at a given effective temperature will be a more massive star that has migrated to its current position on the HR diagram after a comparatively short life compared to dwarf stars of that temperature.  This, somewhat counterintuitively, will generally mean that the evolved star will be younger than the dwarfs at similar temperatures and so will have formed at a metallicity the same or higher than its main sequence counterparts.  However, during a star's main sequence lifetime the surface abundances will tend to decrease due to diffusion and gravitational settling out of the convective layer \citep{1984ApJ...282..206M,1986ApJS...61..177P,2002JGRA..107.1442T,2017ApJ...840...99D}.  Once the star begins to evolve off of the main sequence and its convective region begins to grow, some of this material can be re-mixed into the convective region and the surface abundance will be closer to or exceed its initial value \citep{2017ApJ...840...99D,2018ApJ...857...14S}.  Ratios between most elements remain relatively constant.

\begin{figure}[htbp!] 
   \centering
   \includegraphics[width=\columnwidth]{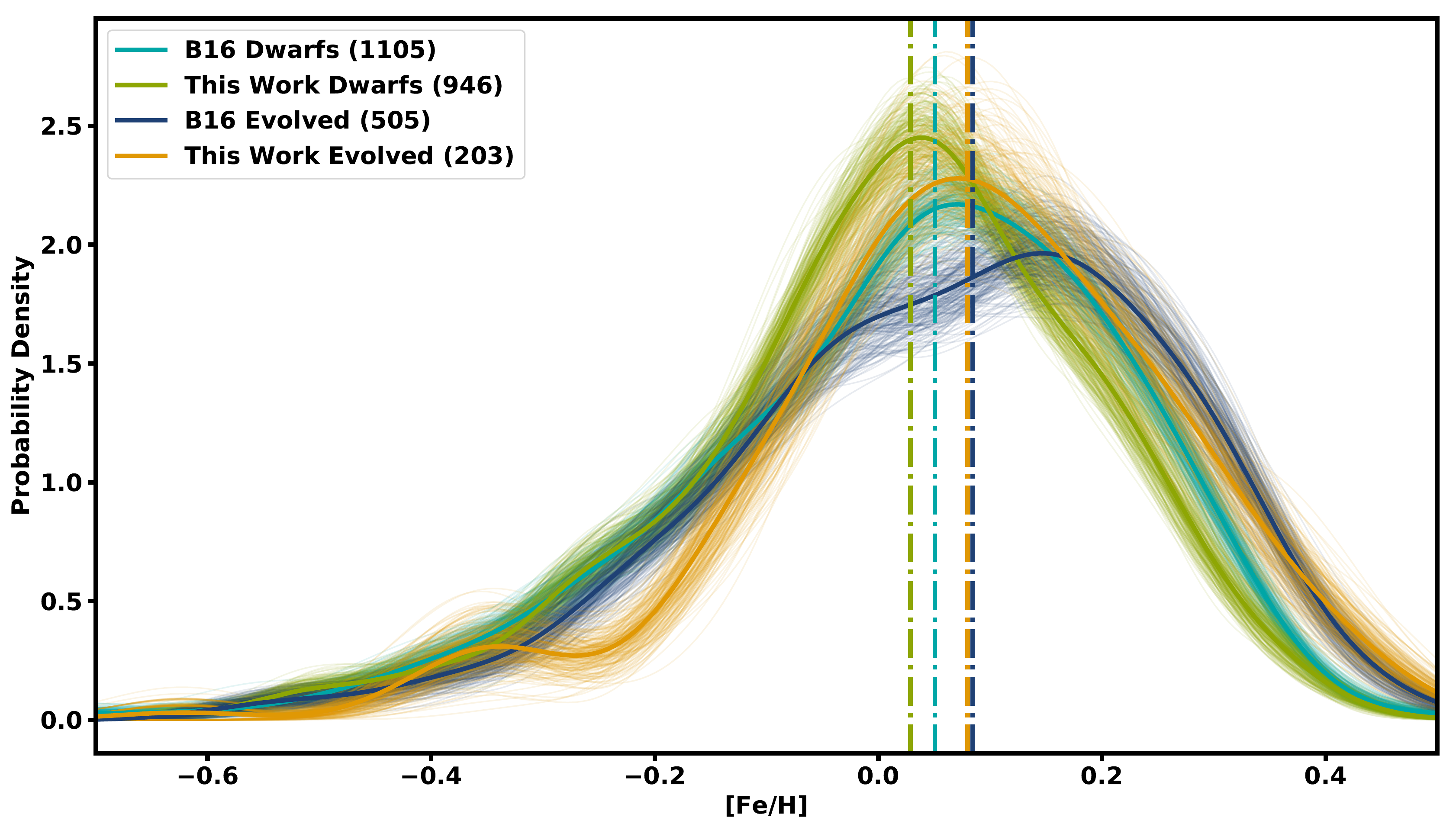} 
   \caption{[Fe/H] distributions for dwarf (teal blue and green lines) and evolved (dark blue and yellow lines) stars in the \citetalias{Brewer:2016gf} sample and this work.  Evolved star abundances are slightly higher overall, conforming to our expectations of higher metallicity in evolved stars due to expanded convective zones. The bold lines are PDFs from a Gaussian KDE fit to the data and the thin lines are the same for 200 bootstrap realizations of each sample. Vertical lines mark the median of each sample.}
   \label{fig:feh_comparison}
\end{figure}

\begin{figure}[htbp!] 
   \centering
   \includegraphics[width=\columnwidth]{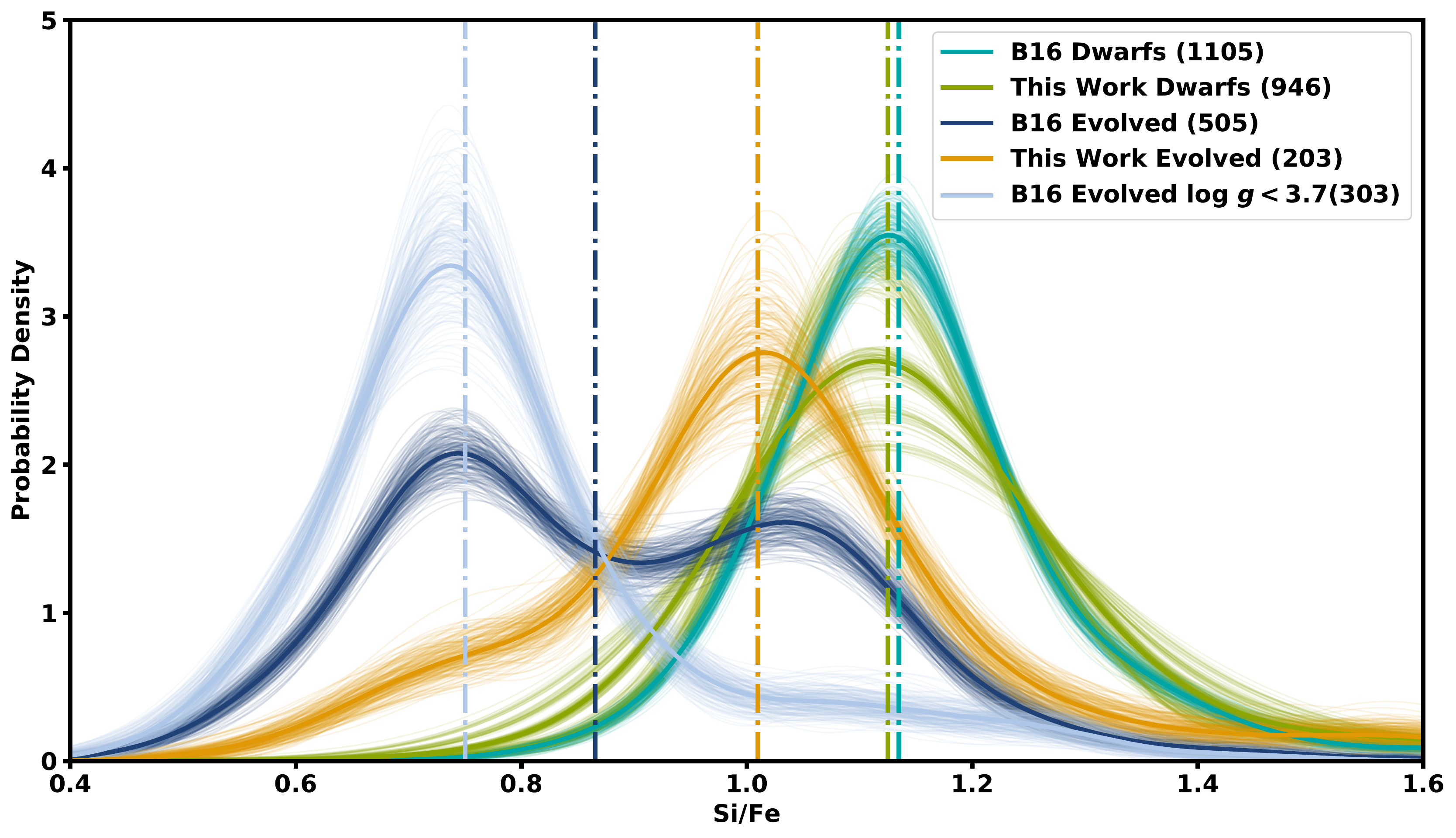} 
   \caption{ Si/Fe distributions for the same sub-samples as in Figure \ref{fig:feh_comparison}.  If the evolved stars and dwarfs started with the same abundances, then the Si/Fe ratio should be similar.  The differences in the distributions points to possible systematic errors in recovering abundances for evolved stars.  We see confirmation of this in the double peaked distribution of evolved stars in \citetalias{Brewer:2016gf} (dark blue lines).  We broke out just those stars with  \logg\ $< 3.7$ in the \citetalias{Brewer:2016gf} evolved star sample (light blue lines) and see that they are almost entirely low Si/Fe stars. The bold lines are PDFs from a Gaussian KDE fit to the data and the thin lines are the same for 200 bootstrap realizations of each sample. Vertical lines mark the median of each sample.}
   \label{fig:sife_comparison}
\end{figure}

\begin{figure*}[htb!] 
   \centering
   \includegraphics[width=\textwidth]{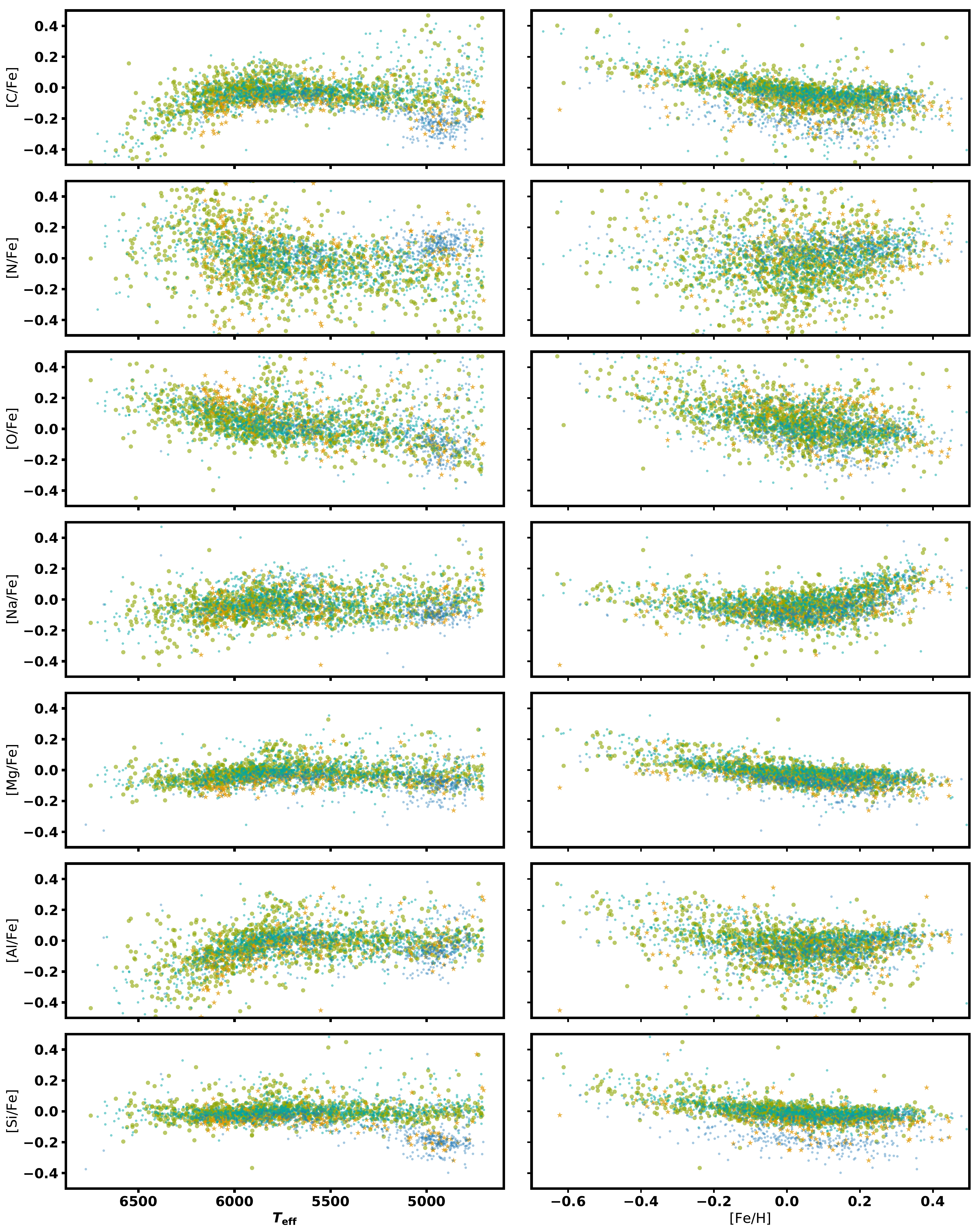} 
   \caption{Light elements [X/Fe], after applying empirical correction from \citetalias{Brewer:2016gf}, plotted against both temperature and [Fe/H].}
   \label{fig:light_elems_x_fe}
\end{figure*}

\begin{figure*}[htb!] 
   \centering
   \includegraphics[width=\textwidth]{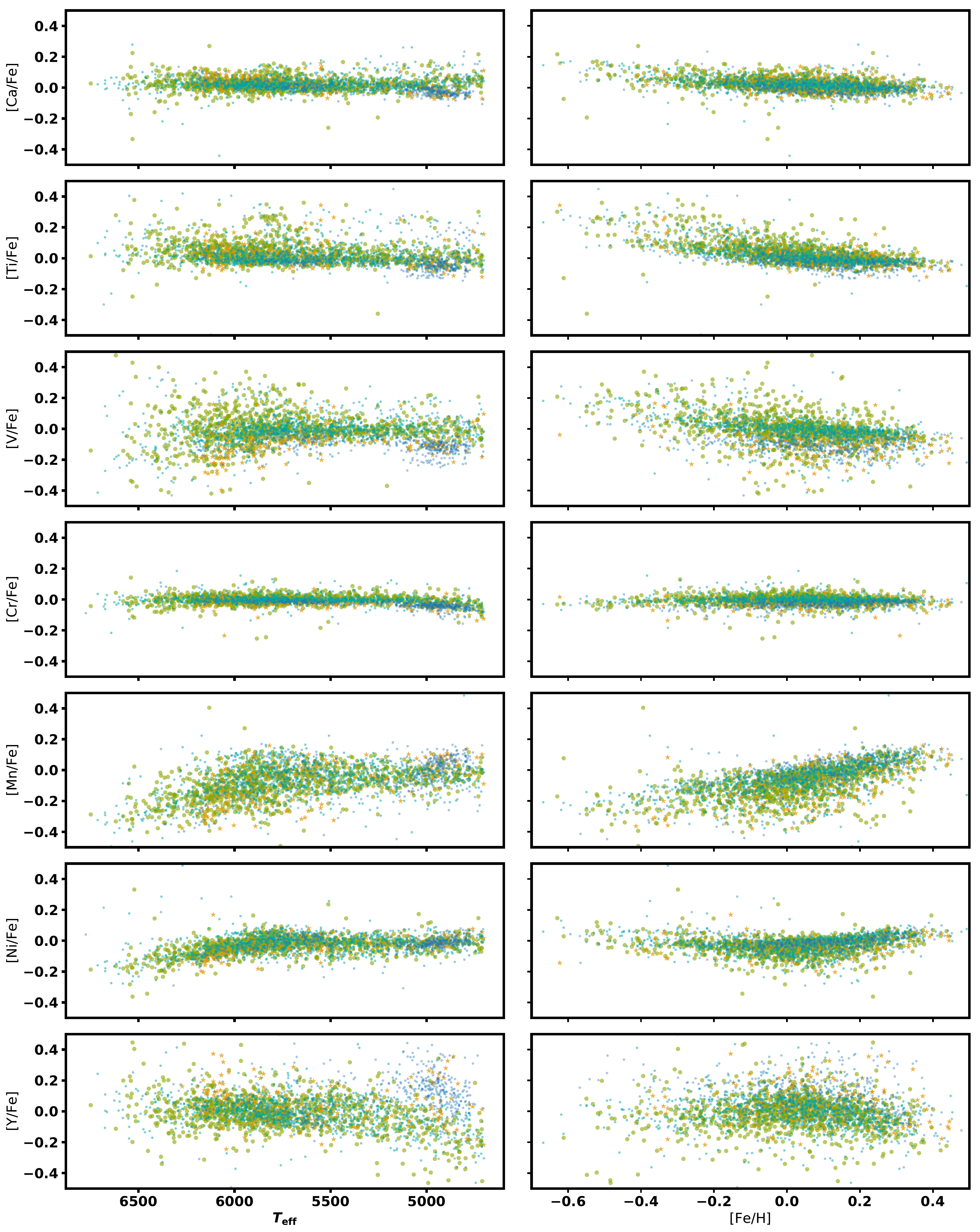} 
   \caption{Heavy elements [X/Fe], after applying empirical correction from \citetalias{Brewer:2016gf},  plotted against both temperature and [Fe/H].}
   \label{fig:heavy_elems_x_fe}
\end{figure*}

In both the \citetalias{Brewer:2016gf} sample and this work,  the dwarf stars are very slightly more metal poor than the evolved stars (Figure \ref{fig:feh_comparison}), which agrees qualitatively with our expectations.  When looking at the [Si/Fe] ratios, which should match more closely, we instead see that the \citetalias{Brewer:2016gf} evolved stars show a significant shift to lower Si/Fe (Figure\ref{fig:sife_comparison}).  Closer examination shows that all of the stars with lower Si/Fe have \logg\ $< 3.7$, placing them as older subgiants, or at the base of the giant branch.  There are comparatively few of these stars in the current work but we do see that all evolved stars are reporting a systematically lower Si/Fe than dwarfs.  This points to a systematic trends in abundance with \logg\ for the evolved stars with lower surface gravities. We can also see some indications of these effects by eye when looking at the [X/Fe] ratios as plotted in Figures \ref{fig:light_elems_x_fe} and \ref{fig:heavy_elems_x_fe}.  We only used dwarf stars in our planet radius comparisons, which do not show these trends.

\subsection{Mg/Si Ratios of Small Planet Hosts}

There is no difference in the Mg/Si ratios between hosts of super-Earth sized planets and hosts of sub-Neptune planets (Figure \ref{fig:mgsi_pdf_hosts}), nor is there any difference in the Mg/Si ratio as a function of planet radius between the two groups of planets (Figure \ref{fig:mgsi_contours_planets}).  This contradicts our hypothesis that varying oxygen uptake in planetary cores, leading to different planet masses, could explain the bimodal small planet populations found in \citet{2017AJ....154..109F}.  This eliminates metallicity as a driver for planet size and shows that planet radius evolution is driven by post-formation sculpting of the planetary atmospheres \citep{2013ApJ...775..105O,2017ApJ...847...29O,2017AJ....154..109F}.  There do seem to be differences in planet multiplicity between the two populations, and there are hints of differences in other abundance ratios, but these are beyond the scope of the current paper and will be explored in future studies.

%
\section{Summary and Conclusions} \label{sec:conclusions}

We present a catalog of accurate stellar parameters and precise abundances for \Nstars\ stars, mainly from the \kepler\ field, spectra for which were originally obtained as part of the California-Kepler Survey \citepalias{2017AJ....154..107P}.  The same procedure was used to analyze the spectra of \citetalias{Brewer:2016gf} and \Nduplicates\ stars in this catalog were previously presented in that work.

Due to the low average S/N ratio of spectra in this catalog, we investigated the impact of low S/N on our analysis method. We used high resolution, high S/N spectra of the Sun and a cool star, degraded to the resolution of the Keck HIRES spectrograph with varying amounts of noise applied.  We used these simulated observations to develop an uncertainty correction for all of our parameters and abundances.  We provide independent uncertainties for all of the stars in this catalog and equations to calculate updated uncertainties for the stars in \citetalias{Brewer:2016gf}.

We find that the abundances of stars in this catalog largely agree with those from \citetalias{Brewer:2016gf}, however note some differences that highlight a possible systematic error in the abundances for evolved stars.

We do not find any difference in the Mg/Si ratios between the hosts of super-Earth sized planets when compared to sub-Neptune hosts.  This suggests that chemical differences in their formation are unlikely to have led to the bimodal distribution in sizes.

\movetabledown=2.2in
\begin{rotatetable*}
\input{table_03_stub.tex}
\end{rotatetable*}
\movetabledown=0in

\movetabledown=1.5in
\begin{rotatetable*}
\input{table_04_stub.tex}
\end{rotatetable*}
\movetabledown=0in

\input{table_05_stub.tex}

\movetabledown=1.7in
\begin{rotatetable*}
\setlength{\tabcolsep}{2.5pt} 
\input{table_06_stub.tex}
\setlength{\tabcolsep}{6pt}
\end{rotatetable*}
\movetabledown=0in

%
\acknowledgments
The authors would like to thank the CKS program for obtaining the bulk of the spectra and the many observers who spent their nights in front of the screens.  We also thank Andrew Howard for early access to the spectra, encouragement, and helpful discussions about the data set. Erik Petigura provided good suggestions on improving the paper. Comments from the anonymous referee also helped in making the paper clearer and our analysis more robust. Data presented herein were obtained at the W. M. Keck Observatory from telescope time allocated to the National Aeronautics and Space Administration through the agency's scientific partnership with the California Institute of Technology and the University of California as well as time from the Yale University TAC. The Observatory was made possible by the generous financial support of the W. M. Keck Foundation. This research has made use of the SIMBAD database and the cross-match service, operated at CDS, Strasbourg, France. This publication makes use of data products from the Two Micron All Sky Survey, which is a joint project of the University of Massachusetts and the Infrared Processing and Analysis Center/California Institute of Technology, funded by the National Aeronautics and Space Administration and the National Science Foundation.

The authors wish to recognize and acknowledge the very significant cultural role and reverence that the summit of Mauna Kea has always had within the indigenous Hawaiian community. We are most fortunate to have the opportunity to conduct observations from this mountain.

JMB and DAF acknowledge support from NASA grant 14-ADAP14-0245.

%
\bibliography{ms}


\end{document}

%% file: table_01.tex
\begin{deluxetable}{lcc}
\tablecaption{Mean standard deviations by parameter for Sun and cool star simulations.  The uncertainties from
\citetalias{Brewer:2016gf} are also included here for convenience in calculating total uncertainties.
\label{table:fit_coeffs}
}
\tablehead{\colhead{Parameter} & \colhead{$\sigma_{100}$} & \colhead{B16 Unc}}
\startdata
	$\log g$ & 0.010 & 0.050 \\
	$T_{\mathrm{eff}}$ [K] & 6.000 & 25.000 \\
	{[M/H]} & 0.005 & 0.010 \\
	$v \sin i$ [km/s] & 0.090 & 0.500 \\
	{[C/H]} & 0.011 & 0.026 \\
	{[N/H]} & 0.032 & 0.042 \\
	{[O/H]} & 0.023 & 0.036 \\
	{[Na/H]} & 0.013 & 0.014 \\
	{[Mg/H]} & 0.008 & 0.012 \\
	{[Al/H]} & 0.016 & 0.028 \\
	{[Si/H]} & 0.009 & 0.008 \\
	{[Ca/H]} & 0.010 & 0.014 \\
	{[Ti/H]} & 0.008 & 0.012 \\
	{[V/H]} & 0.013 & 0.034 \\
	{[Cr/H]} & 0.006 & 0.014 \\
	{[Mn/H]} & 0.011 & 0.020 \\
	{[Fe/H]} & 0.006 & 0.010 \\
	{[Ni/H]} & 0.007 & 0.012 \\
	{[Y/H]} & 0.014 & 0.030 \\
\enddata
\end{deluxetable}

%% file: table_02.tex
\begin{deluxetable*}{lDDDDDDD | DDDDDDDDD}
\tablecaption{Scatter in recovered parameters as a function of S/N.
\label{table:snr_uncertainties}
}
\tabletypesize{\footnotesize}
\tabcolsep=0.1cm
\tablehead{\colhead{ } & \multicolumn{14}{c}{Std dev in Solar case at S/N of:} & \multicolumn{18}{c}{Std dev in cool star at S/N of:} \\
\colhead{Parameter} &\multicolumn2c{20} & \multicolumn2c{30} & \multicolumn2c{40} & \multicolumn2c{50} & \multicolumn2c{60} & \multicolumn2c{80} & \multicolumn2c{100} & \multicolumn2c{20} & \multicolumn2c{30} & \multicolumn2c{40} & \multicolumn2c{50} & \multicolumn2c{60} & \multicolumn2c{70} & \multicolumn2c{80} & \multicolumn2c{90} & \multicolumn2c{100} \\
}
\decimals
\startdata
$T_{\mathrm{{eff}}}$ & 27 & 25 & 195 & 14 & 11 & 10 & 7 & 23 & 16 & 11 & 11 & 5 & 5 & 4 & 5 & 4 \\
$ \log g$ & 0.047 & 0.049 & 0.029 & 0.020 & 0.017 & 0.016 & 0.010 & 0.049 & 0.028 & 0.019 & 0.020 & 0.011 & 0.012 & 0.008 & 0.007 & 0.010 \\
{{[M/H]}} & 0.024 & 0.018 & 0.015 & 0.009 & 0.009 & 0.007 & 0.006 & 0.026 & 0.017 & 0.011 & 0.010 & 0.005 & 0.005 & 0.004 & 0.004 & 0.005 \\
$v\sin{i}$ & 0.555 & 0.529 & 0.415 & 0.338 & 0.183 & 0.223 & 0.173 & 0.000 & 0.000 & 0.000 & 0.000 & 0.000 & 0.000 & 0.000 & 0.000 & 0.000 \\
{[C/H]} & 0.046 & 0.040 & 0.027 & 0.016 & 0.013 & 0.013 & 0.011 & 0.052 & 0.031 & 0.018 & 0.015 & 0.015 & 0.015 & 0.012 & 0.007 & 0.010 \\
{[N/H]} & 0.358 & 0.246 & 0.139 & 0.092 & 0.075 & 0.063 & 0.042 & 0.107 & 0.064 & 0.042 & 0.047 & 0.039 & 0.019 & 0.020 & 0.014 & 0.023 \\
{[O/H]} & 0.092 & 0.062 & 0.043 & 0.036 & 0.034 & 0.023 & 0.019 & 0.113 & 0.072 & 0.050 & 0.042 & 0.034 & 0.032 & 0.026 & 0.023 & 0.027 \\
{[Na/H]} & 0.067 & 0.056 & 0.031 & 0.025 & 0.018 & 0.023 & 0.014 & 0.062 & 0.034 & 0.023 & 0.024 & 0.018 & 0.014 & 0.009 & 0.012 & 0.013 \\
{[Mg/H]} & 0.034 & 0.025 & 0.019 & 0.011 & 0.014 & 0.010 & 0.008 & 0.033 & 0.023 & 0.014 & 0.012 & 0.007 & 0.008 & 0.006 & 0.004 & 0.007 \\
{[Al/H]} & 0.071 & 0.062 & 0.056 & 0.032 & 0.029 & 0.020 & 0.018 & 0.077 & 0.048 & 0.026 & 0.020 & 0.021 & 0.014 & 0.020 & 0.019 & 0.015 \\
{[Si/H]} & 0.041 & 0.025 & 0.021 & 0.020 & 0.014 & 0.013 & 0.009 & 0.035 & 0.019 & 0.016 & 0.016 & 0.009 & 0.012 & 0.009 & 0.010 & 0.009 \\
{[Ca/H]} & 0.044 & 0.035 & 0.025 & 0.017 & 0.013 & 0.011 & 0.011 & 0.048 & 0.030 & 0.020 & 0.015 & 0.009 & 0.011 & 0.009 & 0.007 & 0.009 \\
{[Ti/H]} & 0.055 & 0.030 & 0.025 & 0.025 & 0.016 & 0.015 & 0.011 & 0.041 & 0.029 & 0.016 & 0.018 & 0.006 & 0.008 & 0.007 & 0.008 & 0.006 \\
{[V/H]} & 0.106 & 0.050 & 0.042 & 0.037 & 0.027 & 0.026 & 0.017 & 0.046 & 0.030 & 0.023 & 0.019 & 0.011 & 0.010 & 0.008 & 0.008 & 0.009 \\
{[Cr/H]} & 0.027 & 0.022 & 0.018 & 0.019 & 0.010 & 0.011 & 0.007 & 0.023 & 0.017 & 0.015 & 0.014 & 0.007 & 0.006 & 0.007 & 0.005 & 0.005 \\
{[Mn/H]} & 0.051 & 0.041 & 0.029 & 0.020 & 0.014 & 0.015 & 0.010 & 0.059 & 0.036 & 0.026 & 0.020 & 0.015 & 0.016 & 0.011 & 0.011 & 0.011 \\
{[Fe/H]} & 0.025 & 0.017 & 0.015 & 0.009 & 0.008 & 0.007 & 0.006 & 0.027 & 0.017 & 0.012 & 0.010 & 0.005 & 0.006 & 0.005 & 0.005 & 0.005 \\
{[Ni/H]} & 0.035 & 0.023 & 0.020 & 0.012 & 0.012 & 0.007 & 0.008 & 0.031 & 0.018 & 0.012 & 0.012 & 0.008 & 0.007 & 0.004 & 0.005 & 0.005 \\
{[Y/H]} & 0.062 & 0.032 & 0.039 & 0.028 & 0.018 & 0.018 & 0.015 & 0.066 & 0.039 & 0.030 & 0.026 & 0.020 & 0.017 & 0.014 & 0.012 & 0.014 \\
\enddata
\end{deluxetable*}

%% file: table_03_stub.tex
\begin{deluxetable*}{llllrrrrrrrrrrrrr}
\tablecaption{Spectroscopically determined stellar properties.
\label{table:stellar_properties}
}
\tablehead{\colhead{ID} & \colhead{Name} & \colhead{$\alpha$} & \colhead{$\delta$} & \colhead{$T_{\mathrm{eff}}$} & \colhead{$\log g$} & \colhead{{[M/H]}} & \colhead{S$_{\mathrm{HK}}$} & \colhead{$\log{R_{HK}}$} & \colhead{$v_{{broad}}$} & \colhead{$v\sin{i}$} & \colhead{$v_{\mathrm{{mac}}}$} & \colhead{$v_{\mathrm{{rad}}}$} & \colhead{S/N} & \colhead{C-rms} & \colhead{L-rms} & \colhead{N} \\
\colhead{ } & \colhead{ } & \colhead{J2000} & \colhead{J2000} & \colhead{(K)} & \colhead{(cm s$^{-2}$)} & \colhead{ } & \colhead{ } & \colhead{ } & \colhead{(km s$^{-1}$)} & \colhead{(km s$^{-1}$)} & \colhead{(km s$^{-1}$)} & \colhead{(km s$^{-1}$)} & \colhead{6000 \AA} & \colhead{ } & \colhead{ } & \colhead{ } \\
\colhead{(1)} & \colhead{(2)} & \colhead{(3)} & \colhead{(4)} & \colhead{(5)} & \colhead{(6)} & \colhead{(7)} & \colhead{(8)} & \colhead{(9)} & \colhead{(10)} & \colhead{(11)} & \colhead{(12)} & \colhead{(13)} & \colhead{(14)} & \colhead{(15)} & \colhead{(16)} & \colhead{(17)}}
\startdata
2281 & KOI-3248 & 19 21 51.6 & +48 19 56 & 5742 & 4.34 & -0.07 & 0.09 & nan & 3.7 & 1.4 & 3.4 & -7.1 & 50 & 0.02 & 0.02 & 1 \\
2361 & KOI-4273 & 19 36 50.4 & +46 28 48 & 6123 & 4.11 & 0.14 & 0.13 & -5.22 & 7.3 & 4.5 & 5.6 & -21.3 & 46 & 0.03 & 0.03 & 1 \\
2393 & KOI-3605 & 19 43 21.2 & +41 25 38 & 5252 & 4.39 & 0.05 & 0.14 & -5.18 & 3.1 & 2.1 & 2.1 & -0.1 & 51 & 0.02 & 0.03 & 1 \\
2405 & KOI-3197 & 19 45 09.7 & +44 25 24 & 6218 & 4.43 & -0.02 & 0.17 & -4.87 & 9.3 & 6.1 & 6.4 & 2.2 & 48 & 0.02 & 0.03 & 1 \\
2430 & KOI-1353 & 19 49 51.7 & +42 52 58 & 5951 & 4.46 & -0.02 & 0.25 & -4.60 & 7.3 & 5.5 & 4.4 & -19.1 & 47 & 0.03 & 0.03 & 1 \\
2435 & KOI-4323 & 19 50 19.7 & +40 31 53 & 6314 & 4.22 & -0.03 & 0.06 & nan & 12.9 & 9.4 & 7.4 & -52.7 & 41 & 0.03 & 0.04 & 1 \\
2445 & KOI-3384 & 19 52 22.9 & +44 44 13 & 5994 & 4.24 & 0.07 & 0.11 & -5.51 & 5.4 & 2.7 & 4.6 & -12.1 & 50 & 0.02 & 0.02 & 1 \\
2460 & KOI-4409 & 19 58 08.4 & +40 28 40 & 5872 & 4.26 & 0.21 & 0.17 & -4.94 & 5.4 & 3.5 & 3.9 & -11.9 & 52 & 0.09 & 0.04 & 1 \\
2464 & KOI-4226 & 20 02 15.0 & +46 02 16 & 5817 & 4.02 & 0.37 & 0.10 & -5.89 & 6.1 & 4.7 & 3.7 & -28.2 & 43 & 0.03 & 0.07 & 1 \\
2485 & K2-58 & 22 15 17.2 & -14 03 00 & 5038 & 4.50 & 0.19 & 0.34 & -4.67 & 2.9 & 2.2 & 1.8 & 2.0 & 94 & 0.02 & 0.02 & 1 \\
\enddata
\tablecomments{Table \ref{table:stellar_properties} is published in its entirety in the electronic edition of this article. A portion is shown here for guidance regarding its form and content.}
\end{deluxetable*}

%% file: table_04_stub.tex
\begin{deluxetable*}{llrrrrrrrrrrrrrrr}
\tablecaption{Spectroscopically determined stellar elemental abundances.
\label{table:stellar_abunds}
}
\tablehead{\colhead{ID} & \colhead{NAME} & \colhead{[C/H]} & \colhead{[N/H]} & \colhead{[O/H]} & \colhead{[Na/H]} & \colhead{[Mg/H]} & \colhead{[Al/H]} & \colhead{[Si/H]} & \colhead{[Ca/H]} & \colhead{[Ti/H]} & \colhead{[V/H]} & \colhead{[Cr/H]} & \colhead{[Mn/H]} & \colhead{[Fe/H]} & \colhead{[Ni/H]} & \colhead{[Y/H]} \\
\colhead{(1)} & \colhead{(2)} & \colhead{(18)} & \colhead{(19)} & \colhead{(20)} & \colhead{(21)} & \colhead{(22)} & \colhead{(23)} & \colhead{(24)} & \colhead{(25)} & \colhead{(26)} & \colhead{(27)} & \colhead{(28)} & \colhead{(29)} & \colhead{(30)} & \colhead{(31)} & \colhead{(32)}}
\startdata
2281 & KOI-3248 & -0.06 & -0.04 & -0.05 & -0.17 & -0.07 & 0.03 & -0.04 & -0.06 & -0.06 & -0.16 & -0.06 & -0.12 & -0.08 & -0.11 & -0.15 \\
2361 & KOI-4273 & 0.07 & -0.02 & 0.18 & 0.14 & 0.11 & 0.13 & 0.08 & 0.24 & 0.25 & -0.06 & 0.18 & 0.01 & 0.17 & 0.10 & 0.34 \\
2393 & KOI-3605 & 0.01 & 0.08 & -0.03 & 0.17 & 0.05 & 0.09 & 0.09 & 0.09 & 0.09 & 0.10 & 0.10 & 0.03 & 0.07 & 0.09 & -0.12 \\
2405 & KOI-3197 & -0.03 & -0.04 & -0.02 & -0.08 & -0.01 & -0.21 & -0.04 & 0.03 & 0.00 & 0.03 & 0.03 & -0.20 & 0.00 & -0.09 & 0.15 \\
2430 & KOI-1353 & -0.04 & 0.03 & -0.00 & -0.14 & -0.06 & -0.01 & -0.04 & 0.08 & -0.01 & -0.15 & 0.01 & -0.10 & 0.03 & -0.08 & 0.04 \\
2435 & KOI-4323 & 0.03 & 0.62 & 0.05 & -0.05 & -0.04 & -0.02 & -0.07 & 0.02 & 0.10 & -0.58 & 0.03 & -0.17 & -0.01 & -0.09 & -0.01 \\
2445 & KOI-3384 & 0.01 & 0.16 & 0.04 & 0.08 & 0.06 & 0.08 & 0.07 & 0.12 & 0.05 & -0.03 & 0.07 & -0.01 & 0.08 & 0.08 & 0.05 \\
2460 & KOI-4409 & 0.16 & 0.36 & 0.23 & 0.31 & 0.18 & 0.20 & 0.26 & 0.21 & 0.21 & 0.28 & 0.19 & 0.29 & 0.23 & 0.23 & 0.18 \\
2464 & KOI-4226 & 0.32 & 0.43 & 0.43 & 0.55 & 0.32 & 0.41 & 0.31 & 0.31 & 0.37 & 0.29 & 0.40 & 0.53 & 0.37 & 0.43 & 0.47 \\
2485 & K2-58 & 0.17 & 0.09 & 0.14 & 0.21 & 0.16 & 0.20 & 0.23 & 0.24 & 0.21 & 0.22 & 0.24 & 0.27 & 0.25 & 0.22 & 0.17
\enddata
\tablecomments{Table \ref{table:stellar_abunds} is published in its entirety in the electronic edition of this article. A portion is shown here for guidance regarding its form and content.}
\end{deluxetable*}

%% file: table_05_stub.tex
\begin{deluxetable*}{llDDrrrrrr}
\tablecaption{Derived Stellar Properties.
    \label{table:derived_stellar_props}
    }
\tablehead{\colhead{ID} & \colhead{Name} & \twocolhead{Plx} & \twocolhead{$\sigma$ Plx} & \colhead{R} & \colhead{$\Delta$ R} & \colhead{M} & \colhead{$\Delta$ M} & \colhead{Age} & \colhead{$\Delta$ Age} \\
 \colhead{} & \colhead{} & \twocolhead{(mas)} & \twocolhead{(mas)} & \colhead{(R$_{\odot}$)} & \colhead{(R$_{\odot}$)} & \colhead{(M$_{\odot}$)} & \colhead{(M$_{\odot}$)} & \colhead{(Gyrs)} & \colhead{(Gyrs)} \\
 \colhead{(1)} & \colhead{(2)} & \twocolhead{(33)} & \twocolhead{(34)} & \colhead{(35)} & \colhead{(36)} & \colhead{(37)} & \colhead{(38)} & \colhead{(39)} & \colhead{(40)}}
\decimals
\startdata
2281 & KOI-3248 & 2.7 & 0.023 & 1.14 & 1.12 -  1.16 & 0.96 & 0.95 -  0.97 & 8.40 & 8.01 -  8.73 \\
2361 & KOI-4273 & -1.31 & 0.5 & 1.81 & 1.69 -  1.95 & 1.34 & 1.29 -  1.39 & 3.48 & 3.29 -  3.76 \\
2393 & KOI-3605 & 1.78 & 0.018 & 0.98 & 0.97 -  0.99 & 0.86 & 0.85 -  0.86 & 14.77 & 14.40 - 14.94 \\
2405 & KOI-3197 & 3.54 & 0.035 & 1.12 & 1.10 -  1.13 & 1.16 & 1.15 -  1.18 & 1.08 & 0.61 -  1.60 \\
2430 & KOI-1353 & 1.39 & 0.018 & 1.03 & 1.01 -  1.05 & 1.07 & 1.06 -  1.09 & 2.22 & 1.40 -  3.03 \\
2435 & KOI-4323 & 0.93 & 0.015 & 1.74 & 1.70 -  1.78 & 1.33 & 1.32 -  1.35 & 3.06 & 2.96 -  3.14 \\
2445 & KOI-3384 & 1.6 & 0.017 & 1.29 & 1.27 -  1.32 & 1.11 & 1.10 -  1.12 & 5.20 & 4.89 -  5.54 \\
2460 & KOI-4409 & 2.2 & 0.028 & 1.28 & 1.25 -  1.30 & 1.12 & 1.10 -  1.13 & 5.36 & 4.96 -  5.76 \\
2464 & KOI-4226 & 1.34 & 0.031 & 1.98 & 1.93 -  2.04 & 1.39 & 1.37 -  1.40 & 3.62 & 3.48 -  3.80 \\
2485 & K2-58 & 5.47 & 0.043 &  &  &  &  &  & \\
\enddata
\tablecomments{Table \ref{table:derived_stellar_props} is published in its entirety in the electronic edition of this article. A portion is shown here for guidance regarding its form and content.}
\end{deluxetable*}

%% file: table_06_stub.tex
\begin{deluxetable*}{llrrrrrrrrrrrrrrrrrrr}
\tablecaption{Uncertainties for spectroscopically determined stellar properties and abundances.
\label{table:uncertainties}
}
\tablehead{\colhead{ID} & \colhead{Name} & \colhead{$T_{\mathrm{{eff}}}$} & \colhead{$ \log g$} & \colhead{{{[M/H]}}} & \colhead{$v\sin{i}$} & \colhead{{[C/H]}} & \colhead{{[N/H]}} & \colhead{{[O/H]}} & \colhead{{[Na/H]}} & \colhead{{[Mg/H]}} & \colhead{{[Al/H]}} & \colhead{{[Si/H]}} & \colhead{{[Ca/H]}} & \colhead{{[Ti/H]}} & \colhead{{[V/H]}} & \colhead{{[Cr/H]}} & \colhead{{[Mn/H]}} & \colhead{{[Fe/H]}} & \colhead{{[Ni/H]}} & \colhead{{[Y/H]}} \\
\colhead{(1)} & \colhead{(2)} & \colhead{(3)} & \colhead{(4)} & \colhead{(5)} & \colhead{(6)} & \colhead{(7)} & \colhead{(8)} & \colhead{(9)} & \colhead{(10)} & \colhead{(11)} & \colhead{(12)} & \colhead{(13)} & \colhead{(14)} & \colhead{(15)} & \colhead{(16)} & \colhead{(17)} & \colhead{(18)} & \colhead{(19)} & \colhead{(20)} & \colhead{(21)}}
\startdata
2281 & KOI-3248 & 27 & 0.05 & 0.01 & 0.5 & 0.03 & 0.08 & 0.06 & 0.03 & 0.02 & 0.04 & 0.02 & 0.02 & 0.02 & 0.04 & 0.02 & 0.03 & 0.02 & 0.02 & 0.04 \\
2361 & KOI-4273 & 28 & 0.05 & 0.01 & 0.5 & 0.04 & 0.08 & 0.06 & 0.03 & 0.02 & 0.04 & 0.02 & 0.03 & 0.02 & 0.04 & 0.02 & 0.03 & 0.02 & 0.02 & 0.04 \\
2393 & KOI-3605 & 27 & 0.05 & 0.01 & 0.5 & 0.03 & 0.07 & 0.06 & 0.03 & 0.02 & 0.04 & 0.02 & 0.02 & 0.02 & 0.04 & 0.02 & 0.03 & 0.02 & 0.02 & 0.04 \\
2405 & KOI-3197 & 27 & 0.05 & 0.01 & 0.5 & 0.03 & 0.08 & 0.06 & 0.03 & 0.02 & 0.04 & 0.02 & 0.03 & 0.02 & 0.04 & 0.02 & 0.03 & 0.02 & 0.02 & 0.04 \\
2430 & KOI-1353 & 28 & 0.05 & 0.01 & 0.5 & 0.03 & 0.08 & 0.06 & 0.03 & 0.02 & 0.04 & 0.02 & 0.03 & 0.02 & 0.04 & 0.02 & 0.03 & 0.02 & 0.02 & 0.04 \\
2435 & KOI-4323 & 28 & 0.06 & 0.02 & 0.5 & 0.04 & 0.09 & 0.07 & 0.03 & 0.02 & 0.05 & 0.02 & 0.03 & 0.02 & 0.05 & 0.02 & 0.03 & 0.02 & 0.02 & 0.04 \\
2445 & KOI-3384 & 27 & 0.05 & 0.01 & 0.5 & 0.03 & 0.08 & 0.06 & 0.03 & 0.02 & 0.04 & 0.02 & 0.02 & 0.02 & 0.04 & 0.02 & 0.03 & 0.02 & 0.02 & 0.04 \\
2460 & KOI-4409 & 27 & 0.05 & 0.01 & 0.5 & 0.03 & 0.07 & 0.06 & 0.03 & 0.02 & 0.04 & 0.02 & 0.02 & 0.02 & 0.04 & 0.02 & 0.03 & 0.02 & 0.02 & 0.04 \\
2464 & KOI-4226 & 28 & 0.06 & 0.02 & 0.5 & 0.04 & 0.09 & 0.06 & 0.03 & 0.02 & 0.05 & 0.02 & 0.03 & 0.02 & 0.05 & 0.02 & 0.03 & 0.02 & 0.02 & 0.04 \\
2485 & K2-58 & 25 & 0.05 & 0.01 & 0.5 & 0.03 & 0.05 & 0.04 & 0.02 & 0.01 & 0.03 & 0.01 & 0.02 & 0.01 & 0.04 & 0.02 & 0.02 & 0.01 & 0.01 & 0.03 \\
\enddata
\tablecomments{Table \ref{table:uncertainties} is published in its entirety in the electronic edition of this article. A portion is shown here for guidance regarding its form and content.}
\end{deluxetable*}

%% file: ms.bbl
\begin{thebibliography}{}
\expandafter\ifx\csname natexlab\endcsname\relax\def\natexlab#1{#1}\fi

\bibitem[{Asplund {et~al.}(2009)Asplund, Grevesse, Sauval, \&
  Scott}]{2009ARA&A..47..481A}
Asplund, M., Grevesse, N., Sauval, A.~J., \& Scott, P. 2009, Annual Review of
  Astronomy {\&} Astrophysics, 47, 481

\bibitem[{Bergemann {et~al.}(2012)Bergemann, Lind, Collet, Magic, \&
  Asplund}]{2012MNRAS.427...27B}
Bergemann, M., Lind, K., Collet, R., Magic, Z., \& Asplund, M. 2012, Monthly
  Notices of the Royal Astronomical Society, 427, 27

\bibitem[{Borucki {et~al.}(2011)Borucki, Koch, Basri, Batalha, Brown, Bryson,
  Caldwell, Christensen-Dalsgaard, Cochran, Devore, Dunham, Gautier, Geary,
  Gilliland, Gould, Howell, Jenkins, Latham, Lissauer, Marcy, Rowe, Sasselov,
  Boss, Charbonneau, Ciardi, Doyle, Dupree, Ford, Fortney, Holman, Seager,
  Steffen, Tarter, Welsh, Allen, Buchhave, Christiansen, Clarke, Das, Desert,
  Endl, Fabrycky, Fressin, Haas, Horch, Howard, Isaacson, Kjeldsen,
  Kolodziejczak, Kulesa, Li, Lucas, Machalek, McCarthy, MacQueen, Meibom,
  Miquel, Prsa, Quinn, Quintana, Ragozzine, Sherry, Shporer, Tenenbaum, Torres,
  Twicken, Van~Cleve, Walkowicz, Witteborn, \& Still}]{2011ApJ...736...19B}
Borucki, W.~J., Koch, D.~G., Basri, G., {et~al.} 2011, The Astrophysical
  Journal, 736, 19

\bibitem[{Brewer {et~al.}(2015)Brewer, Fischer, Basu, Valenti, \&
  Piskunov}]{2015ApJ...805..126B}
Brewer, J.~M., Fischer, D.~A., Basu, S., Valenti, J.~A., \& Piskunov, N. 2015,
  The Astrophysical Journal, 805, 126

\bibitem[{Brewer {et~al.}(2016)Brewer, Fischer, Valenti, \&
  Piskunov}]{Brewer:2016gf}
Brewer, J.~M., Fischer, D.~A., Valenti, J.~A., \& Piskunov, N. 2016, The
  Astrophysical Journal Supplement Series, 225, 32

\bibitem[{Brugamyer {et~al.}(2011)Brugamyer, Dodson-Robinson, Cochran, \&
  Sneden}]{2011ApJ...738...97B}
Brugamyer, E., Dodson-Robinson, S.~E., Cochran, W.~D., \& Sneden, C. 2011, The
  Astrophysical Journal, 738, 97

\bibitem[{Buchhave {et~al.}(2014)Buchhave, Bizzarro, Latham, Sasselov, Cochran,
  Endl, Isaacson, Juncher, \& Marcy}]{2014Natur.509..593B}
Buchhave, L.~A., Bizzarro, M., Latham, D.~W., {et~al.} 2014, Nature, 509, 593

\bibitem[{Burke {et~al.}(2015)Burke, Christiansen, Mullally, Seader, Huber,
  Rowe, Coughlin, Thompson, Catanzarite, Clarke, Morton, Caldwell, Bryson,
  Haas, Batalha, Jenkins, Tenenbaum, Twicken, Li, Quintana, Barclay, Henze,
  Borucki, Howell, \& Still}]{2015ApJ...809....8B}
Burke, C.~J., Christiansen, J.~L., Mullally, F., {et~al.} 2015, The
  Astrophysical Journal, 809, 8

\bibitem[{Collaboration {et~al.}(2016)Collaboration, Prusti, de~Bruijne, Brown,
  Vallenari, Babusiaux, Bailer-Jones, Bastian, Biermann, Evans, Eyer, Jansen,
  Jordi, Klioner, Lammers, Lindegren, Luri, Mignard, Milligan, Panem,
  Poinsignon, Pourbaix, Randich, Sarri, Sartoretti, Siddiqui, Soubiran,
  Valette, van Leeuwen, Walton, Aerts, Arenou, Cropper, Drimmel, H{\o}g, Katz,
  Lattanzi, O'Mullane, Grebel, Holland, Huc, Passot, Bramante, Cacciari,
  Casta{\~n}eda, Chaoul, Cheek, De~Angeli, Fabricius, Guerra, Hern{\'a}ndez,
  Jean-Antoine-Piccolo, Masana, Messineo, Mowlavi, Nienartowicz,
  Ord{\'o}{\~n}ez-Blanco, Panuzzo, Portell, Richards, Riello, Seabroke, Tanga,
  Th{\'e}venin, Torra, Els, Gracia-Abril, Comoretto, Garcia-Reinaldos, Lock,
  Mercier, Altmann, Andrae, Astraatmadja, Bellas-Velidis, Benson, Berthier,
  Blomme, Busso, Carry, Cellino, Clementini, Cowell, Creevey, Cuypers,
  Davidson, De~Ridder, de~Torres, Delchambre, Dell'Oro, Ducourant, Fr{\'e}mat,
  Garc{\'\i}a-Torres, Gosset, Halbwachs, Hambly, Harrison, Hauser, Hestroffer,
  Hodgkin, Huckle, Hutton, Jasniewicz, Jordan, Kontizas, Korn, Lanzafame,
  Manteiga, Moitinho, Muinonen, Osinde, Pancino, Pauwels, PETIT, Recio-Blanco,
  Robin, Sarro, Siopis, Smith, Smith, Sozzetti, Thuillot, van Reeven, Viala,
  Abbas, Abreu~Aramburu, Accart, Aguado, Allan, Allasia, Altavilla,
  {\'A}lvarez, Alves, Anderson, Andrei, Anglada~Varela, Antiche, Antoja, Anton,
  Arcay, Atzei, Ayache, Bach, Baker, Balaguer-N{\'u}{\~n}ez, Barache, Barata,
  Barbier, Barblan, Baroni, Barrado~y Navascu{\'e}s, Barros, Barstow, Becciani,
  Bellazzini, Bellei, Bello~Garc{\'\i}a, Belokurov, Bendjoya, Berihuete,
  Bianchi, Bienaym{\'e}, Billebaud, Blagorodnova, Blanco-Cuaresma, Boch,
  Bombrun, Borrachero, Bouquillon, Bourda, Bouy, Bragaglia, Breddels,
  Brouillet, Br{\"u}semeister, Bucciarelli, Budnik, Burgess, Burgon, Burlacu,
  Busonero, Buzzi, Caffau, Cambras, Campbell, Cancelliere, Cantat-Gaudin,
  Carlucci, Carrasco, Castellani, Charlot, Charnas, Charvet, Chassat,
  Chiavassa, Clotet, Cocozza, Collins, Collins, Costigan, Crifo, Cross, Crosta,
  Crowley, Dafonte, Damerdji, Dapergolas, David, David, De~Cat, de~Felice,
  de~Laverny, De~Luise, De~March, de~Martino, de~Souza, Debosscher, del Pozo,
  Delbo, Delgado, Delgado, di~Marco, Di~Matteo, Diakite, Distefano, Dolding,
  Dos~Anjos, Drazinos, Duran, Dzigan, Ecale, Edvardsson, Enke, Erdmann,
  Escolar, Espina, Evans, Eynard~Bontemps, Fabre, Fabrizio, Faigler,
  Falc{\~a}o, Farr{\`a}s~Casas, Faye, Federici, Fedorets,
  Fern{\'a}ndez-Hern{\'a}ndez, Fernique, Fienga, Figueras, Filippi, Findeisen,
  Fonti, Fouesneau, Fraile, Fraser, Fuchs, Furnell, Gai, Galleti, Galluccio,
  Garabato, Garc{\'\i}a-Sedano, Gar{\'e}, Garofalo, Garralda, Gavras, Gerssen,
  Geyer, Gilmore, Girona, Giuffrida, Gomes, Gonz{\'a}lez-Marcos,
  Gonz{\'a}lez-N{\'u}{\~n}ez, Gonz{\'a}lez-Vidal, Granvik, Guerrier, Guillout,
  Guiraud, G{\'u}rpide, Guti{\'e}rrez-S{\'a}nchez, {Guy, L. P.}, Haigron,
  Hatzidimitriou, Haywood, Heiter, Helmi, \& Hobbs}]{2016A&A...595A...1G}
Collaboration, G., Prusti, T., de~Bruijne, J. H.~J., {et~al.} 2016, Astronomy
  and Astrophysics, 595, A1

\bibitem[{Dodson-Robinson {et~al.}(2009)Dodson-Robinson, Willacy, Bodenheimer,
  Turner, \& Beichman}]{2009Icar..200..672D}
Dodson-Robinson, S.~E., Willacy, K., Bodenheimer, P., Turner, N.~J., \&
  Beichman, C.~A. 2009, Icarus, 200, 672

\bibitem[{Dotter {et~al.}(2008)Dotter, Chaboyer, Jevremovi{\'c}, Kostov, Baron,
  \& Ferguson}]{2008ApJS..178...89D}
Dotter, A., Chaboyer, B., Jevremovi{\'c}, D., {et~al.} 2008, The Astrophysical
  Journal Supplement Series, 178, 89

\bibitem[{Dotter {et~al.}(2017)Dotter, Conroy, Cargile, \&
  Asplund}]{2017ApJ...840...99D}
Dotter, A., Conroy, C., Cargile, P., \& Asplund, M. 2017, The Astrophysical
  Journal, 840, 99

\bibitem[{Fischer \& Valenti(2005)}]{2005ApJ...622.1102F}
Fischer, D.~A., \& Valenti, J. 2005, The Astrophysical Journal, 622, 1102

\bibitem[{Fulton {et~al.}(2017)Fulton, Petigura, Howard, Isaacson, Marcy,
  Cargile, Hebb, Weiss, Johnson, Morton, Sinukoff, Crossfield, \&
  Hirsch}]{2017AJ....154..109F}
Fulton, B.~J., Petigura, E.~A., Howard, A.~W., {et~al.} 2017, The Astronomical
  Journal, 154, 109

\bibitem[{Grasset {et~al.}(2009)Grasset, Schneider, \&
  Sotin}]{2009ApJ...693..722G}
Grasset, O., Schneider, J., \& Sotin, C. 2009, The Astrophysical Journal, 693,
  722

\bibitem[{Grevesse {et~al.}(2007)Grevesse, Asplund, \&
  Sauval}]{Grevesse:2007cx}
Grevesse, N., Asplund, M., \& Sauval, A.~J. 2007, Space Science Reviews, Volume
  130, Issue 1-4, pp. 105-114, 130, 105

\bibitem[{Hinkel {et~al.}(2016)Hinkel, Young, Pagano, Desch, Anbar, Adibekyan,
  Blanco-Cuaresma, Carlberg, Mena, Liu, Nordlander, Sousa, Korn, Gruyters,
  Heiter, Jofre, Santos, \& Soubiran}]{2016ApJS..226....4H}
Hinkel, N.~R., Young, P.~A., Pagano, M.~D., {et~al.} 2016, Astrophysical
  Journal, Supplement Series, 226, 4

\bibitem[{Howard {et~al.}(2012)Howard, Marcy, Bryson, Jenkins, Rowe, Batalha,
  Borucki, Koch, Dunham, Gautier, Van~Cleve, Cochran, Latham, Lissauer, Torres,
  Brown, Gilliland, Buchhave, Caldwell, Christensen-Dalsgaard, Ciardi, Fressin,
  Haas, Howell, Kjeldsen, Seager, Rogers, Sasselov, Steffen, Basri,
  Charbonneau, Christiansen, Clarke, Dupree, Fabrycky, Fischer, Ford, Fortney,
  Tarter, Girouard, Holman, Johnson, Klaus, Machalek, Moorhead, Morehead,
  Ragozzine, Tenenbaum, Twicken, Quinn, Isaacson, Shporer, Lucas, Walkowicz,
  Welsh, Boss, Devore, Gould, Smith, Morris, Prsa, Morton, Still, Thompson,
  Mullally, Endl, \& MacQueen}]{2012ApJS..201...15H}
Howard, A.~W., Marcy, G.~W., Bryson, S.~T., {et~al.} 2012, The Astrophysical
  Journal Supplement, 201, 15

\bibitem[{Huitson {et~al.}(2017)Huitson, D{\'e}sert, Bean, Fortney, Stevenson,
  \& Bergmann}]{2017AJ....154...95H}
Huitson, C.~M., D{\'e}sert, J.~M., Bean, J.~L., {et~al.} 2017, The Astronomical
  Journal, 154, 95

\bibitem[{Isaacson \& Fischer(2010)}]{Isaacson:2010gk}
Isaacson, H., \& Fischer, D. 2010, The Astrophysical Journal, 725, 875

\bibitem[{Johnson {et~al.}(2017)Johnson, Petigura, Fulton, Marcy, Howard,
  Isaacson, Hebb, Cargile, Morton, Weiss, Winn, Rogers, Sinukoff, \&
  Hirsch}]{2017AJ....154..108J}
Johnson, J.~A., Petigura, E.~A., Fulton, B.~J., {et~al.} 2017, The Astronomical
  Journal, 154, 108

\bibitem[{Kreidberg {et~al.}(2015)Kreidberg, Line, Bean, Stevenson, Desert,
  Madhusudhan, Fortney, Barstow, Henry, Williamson, \&
  Showman}]{Kreidberg:2015er}
Kreidberg, L., Line, M.~R., Bean, J.~L., {et~al.} 2015, The Astrophysical
  Journal, 814, 66

\bibitem[{Lopez \& Fortney(2013)}]{2013ApJ...776....2L}
Lopez, E.~D., \& Fortney, J.~J. 2013, The Astrophysical Journal, 776, 2

\bibitem[{Lundkvist {et~al.}(2016)Lundkvist, Kjeldsen, Albrecht, Davies, Basu,
  Huber, Justesen, Karoff, Silva~Aguirre, Van~Eylen, Vang, Arentoft, Barclay,
  Bedding, Campante, Chaplin, Christensen-Dalsgaard, Elsworth, Gilliland,
  Handberg, Hekker, Kawaler, Lund, Metcalfe, Miglio, Rowe, Stello, Tingley, \&
  White}]{Lundkvist:2016gf}
Lundkvist, M.~S., Kjeldsen, H., Albrecht, S., {et~al.} 2016, Nature
  Communications, 7, 11201

\bibitem[{Luri {et~al.}(2018)Luri, Brown, Sarro, Arenou, Bailer-Jones,
  Castro-Ginard, de~Bruijne, Prusti, Babusiaux, \&
  Delgado}]{2018arXiv180409376L}
Luri, X., Brown, A. G.~A., Sarro, L.~M., {et~al.} 2018, arXiv.org,
  arXiv:1804.09376

\bibitem[{MacDonald \& Madhusudhan(2017)}]{2017MNRAS.469.1979M}
MacDonald, R.~J., \& Madhusudhan, N. 2017, Monthly Notices of the Royal
  Astronomical Society, 469, 1979

\bibitem[{Michaud {et~al.}(1984)Michaud, Fontaine, \&
  Beaudet}]{1984ApJ...282..206M}
Michaud, G., Fontaine, G., \& Beaudet, G. 1984, Astrophysical Journal, 282, 206

\bibitem[{Morley {et~al.}(2017)Morley, Knutson, Line, Fortney, Thorngren,
  Marley, Teal, \& Lupu}]{Morley:2016vc}
Morley, C.~V., Knutson, H., Line, M., {et~al.} 2017, Astronomical Journal, 153,
  86

\bibitem[{Morton(2015)}]{2015ascl.soft03010M}
Morton, T.~D. 2015, Astrophysics Source Code Library, ascl:1503.010

\bibitem[{Mulders {et~al.}(2016)Mulders, Pascucci, Apai, Frasca, \&
  Molenda-Zakowicz}]{2016AJ....152..187M}
Mulders, G.~D., Pascucci, I., Apai, D., Frasca, A., \& Molenda-Zakowicz, J.
  2016, The Astronomical Journal, 152, 187

\bibitem[{{\"O}berg \& Bergin(2016)}]{2016ApJ...831L..19O}
{\"O}berg, K.~I., \& Bergin, E.~A. 2016, The Astrophysical Journal Letters,
  831, L19

\bibitem[{Owen \& Wu(2013)}]{2013ApJ...775..105O}
Owen, J.~E., \& Wu, Y. 2013, The Astrophysical Journal, 775, 105

\bibitem[{Owen \& Wu(2017)}]{2017ApJ...847...29O}
---. 2017, The Astrophysical Journal, 847, 29

\bibitem[{Paquette {et~al.}(1986)Paquette, Pelletier, Fontaine, \&
  Michaud}]{1986ApJS...61..177P}
Paquette, C., Pelletier, C., Fontaine, G., \& Michaud, G. 1986, Astrophysical
  Journal Supplement Series (ISSN 0067-0049), 61, 177

\bibitem[{Petigura {et~al.}(2017)Petigura, Howard, Marcy, Johnson, Isaacson,
  Cargile, Hebb, Fulton, Weiss, Morton, Winn, Rogers, Sinukoff, Hirsch, \&
  Crossfield}]{2017AJ....154..107P}
Petigura, E.~A., Howard, A.~W., Marcy, G.~W., {et~al.} 2017, The Astronomical
  Journal, 154, 107

\bibitem[{Piskunov \& Valenti(2017)}]{2017A&A...597A..16P}
Piskunov, N., \& Valenti, J.~A. 2017, Astronomy and Astrophysics, 597, A16

\bibitem[{Piso {et~al.}(2016)Piso, Pegues, \& {\"O}berg}]{2016ApJ...833..203P}
Piso, A.-M.~A., Pegues, J., \& {\"O}berg, K.~I. 2016, The Astrophysical
  Journal, 833, 203

\bibitem[{Robinson {et~al.}(2006)Robinson, Laughlin, Bodenheimer, \&
  Fischer}]{Robinson:2006cr}
Robinson, S.~E., Laughlin, G., Bodenheimer, P., \& Fischer, D. 2006, The
  Astrophysical Journal, 643, 484

\bibitem[{Santos {et~al.}(2004)Santos, Israelian, \&
  Mayor}]{2004A&A...415.1153S}
Santos, N.~C., Israelian, G., \& Mayor, M. 2004, Astronomy and Astrophysics,
  415, 1153

\bibitem[{Seager {et~al.}(2007)Seager, Kuchner, Hier~Majumder, \&
  Militzer}]{2007ApJ...669.1279S}
Seager, S., Kuchner, M., Hier~Majumder, C.~A., \& Militzer, B. 2007, The
  Astrophysical Journal, 669, 1279

\bibitem[{Skrutskie {et~al.}(2006)Skrutskie, Cutri, Stiening, Weinberg,
  Schneider, Carpenter, Beichman, Capps, Chester, Elias, Huchra, Liebert,
  Lonsdale, Monet, Price, Seitzer, Jarrett, Kirkpatrick, Gizis, Howard, Evans,
  Fowler, Fullmer, Hurt, Light, Kopan, Marsh, McCallon, Tam, van Dyk, \&
  Wheelock}]{Skrutskie:2006hl}
Skrutskie, M.~F., Cutri, R.~M., Stiening, R., {et~al.} 2006, The Astronomical
  Journal, 131, 1163

\bibitem[{Souto {et~al.}(2018)Souto, Cunha, Smith, Allende~Prieto,
  Garc{\'\i}a-Hern{\'a}ndez, Pinsonneault, Holzer, Frinchaboy, Holtzman,
  Johnson, Jonsson, Majewski, Shetrone, Sobeck, Stringfellow, Teske, Zamora,
  Zasowski, Carrera, Stassun, Fernandez-Trincado, Villanova, Minniti, \&
  Santana}]{2018ApJ...857...14S}
Souto, D., Cunha, K., Smith, V.~V., {et~al.} 2018, The Astrophysical Journal,
  857, 14

\bibitem[{Turcotte \& Wimmer~Schweingruber(2002)}]{2002JGRA..107.1442T}
Turcotte, S., \& Wimmer~Schweingruber, R.~F. 2002, Journal of Geophysical
  Research (Space Physics), 107, 1442

\bibitem[{Unterborn \& Panero(2017)}]{2017ApJ...845...61U}
Unterborn, C.~T., \& Panero, W.~R. 2017, The Astrophysical Journal, 845, 61

\bibitem[{Valenti \& Fischer(2005)}]{2005ApJS..159..141V}
Valenti, J.~A., \& Fischer, D.~A. 2005, The Astrophysical Journal Supplement
  Series, 159, 141

\bibitem[{Wallace {et~al.}(2011)Wallace, Hinkle, Livingston, \&
  Davis}]{2011ApJS..195....6W}
Wallace, L., Hinkle, K.~H., Livingston, W.~C., \& Davis, S.~P. 2011, The
  Astrophysical Journal Supplement, 195, 6

\bibitem[{Wang \& Fischer(2015)}]{2015AJ....149...14W}
Wang, J., \& Fischer, D.~A. 2015, The Astronomical Journal, 149, 14

\bibitem[{Weiss {et~al.}(2018)Weiss, Marcy, Petigura, Fulton, Howard, Winn,
  Isaacson, Morton, Hirsch, Sinukoff, Cumming, Hebb, \&
  Cargile}]{2018AJ....155...48W}
Weiss, L.~M., Marcy, G.~W., Petigura, E.~A., {et~al.} 2018, The Astronomical
  Journal, 155, 48

\bibitem[{Wenger {et~al.}(2000)Wenger, Ochsenbein, Egret, Dubois, Bonnarel,
  Borde, Genova, Jasniewicz, Lalo, Lesteven, \& Monier}]{Wenger:2000ef}
Wenger, M., Ochsenbein, F., Egret, D., {et~al.} 2000, Astronomy and
  Astrophysics Supplement Series, 143, 9

\bibitem[{Winn \& Fabrycky(2015)}]{2015ARA&A..53..409W}
Winn, J.~N., \& Fabrycky, D.~C. 2015, Annual Review of Astronomy {\&}
  Astrophysics, 53, 409

\bibitem[{Zeng {et~al.}(2016)Zeng, Sasselov, \& Jacobsen}]{Zeng:2016cz}
Zeng, L., Sasselov, D.~D., \& Jacobsen, S.~B. 2016, The Astrophysical Journal,
  819, 127

\end{thebibliography}
